\documentclass[a4paper,11pt, fleqn]{article} 
\usepackage[english]{babel}
\usepackage[ansinew]{inputenc}
\usepackage[T1]{fontenc}

\usepackage{palatino}  
\usepackage{caption}
\usepackage{wrapfig}
\usepackage{graphicx}
\usepackage{subfigure}
\usepackage{tabularx}
\usepackage{acronym}
\usepackage{framed}
\usepackage{url}
\usepackage{amsmath}
\usepackage{float}
\usepackage{amsmath}
\usepackage{amssymb}
\usepackage{slashed}
\usepackage{listings}
\usepackage{color}
\usepackage{textcomp}
\usepackage{ dsfont }
\usepackage[usenames,dvipsnames]{xcolor}
\usepackage{titlesec}
\usepackage{wasysym}

\definecolor{orange}{HTML}{FF7F00}
\definecolor{DarkGray}{HTML}{404040}
\definecolor{lightGray}{HTML}{d0d1d1}
\definecolor{hhuBlue}{HTML}{315ba5}

\usepackage{pdfpages}
\usepackage{geometry}
\usepackage{layout}

\date{}

\usepackage{jheppub}
\usepackage{changes}
\usepackage{cancel}

\newcommand{\eqn}[1]{ \begin{equation} #1 \end{equation} }
\newcommand{\eqnsplit}[1]{ \begin{equation}\begin{split} #1 \end{split}\end{equation} }

\newcommand{\lc}{\left (}
\newcommand{\rc}{\right )}
\newcommand{\lL}{\left [}
\newcommand{\rL}{\right ]}

\newcommand{\gA}{\mathcal{A}}

\newcommand{\urlb}[1]{\underline{\url{#1}}}

{\selectlanguage{ngerman}}
{\\\hspace*{\fill}\\\hspace*{\fill}\\}

\newlength{\leftbarwidth}
\newlength{\leftbarsep}
\colorlet{leftbarcolor}{black}

\newcommand*\pFqskip{18mu}
\catcode`,\active
\newcommand*\pFq{\begingroup
        \catcode`\,\active
        \def ,{\mskip\pFqskip\relax}%
        \dopFq
}
\catcode`\,12
\def\dopFq#1#2#3#4#5{%
        {}_{#1}\mathcal{F}_{#2}\lc \left\{#3\right\} ; \left\{#4\right\} ; #5 \rc%
        \endgroup
}

 \title{Nonlinear lepton-photon interactions in external background fields}

\author[a]{Ibrahim Akal}
\author[a,b]{, Gudrid Moortgat-Pick}
\affiliation[a]{Theory Group, Deutsches Elektronen-Synchrotron DESY, Notkestra{\ss}e 85, D-22607 Hamburg, Germany} 
\affiliation[b]{II. Institute for Theoretical Physics, University of Hamburg, Luruper Chaussee 149, D-22761 Hamburg, Germany}

\emailAdd{ibrahim.akal@desy.de}
\emailAdd{gudrid.moortgat-pick@desy.de}
\abstract{
Nonlinear phenomena of lepton-photon interactions in external backgrounds with a generalised periodic plane-wave geometry are studied. 
We discuss nonlinear Compton scattering in head-on lepton-photon collisions extended properly to beyond the soft-photon regime. In addition, our results are applied to stimulated lepton-antilepton pair production in photon collisions with unrestricted energies.
Derivations are considered semi-classically based on unperturbed fermionic Volkov representations encoding the full interaction with the background field. Closed expressions for total probabilities considering S-matrix elements have been derived.
The general formula is applied to Compton scattering by an electron propagating in an external laser-like background. We obtain additive contributions in the extended unconstrained result which turns out to be stringently required in the highly nonlinear regime. A detailed comparison of contributing harmonics is discussed for various field parameters. 
}

\preprint{
\begin{flushright}
DESY 15-180\\
\today
\end{flushright}
}

\begin{document}

\maketitle

\newpage
\section{Introduction}
Quantum electrodynamics (QED) is the fundamental concept of describing relativistic electromagnetic interactions of elementary particles.
The theory is in excellent experimental agreement: the theoretical prediction for the electromagnetic fine structure constant is based on a four-loop calculation of the anomalous magnetic moment of the electron \cite{kinoshita-06, aoyama-07} and is consistent with the measured value on the level of 10 significant digits. 
However, in the presence of strong external background fields, QED acquires novel non-perturbative features and one enters a new interesting field of physics: the strong-field QED (SFQED) \cite{dunne-09, heinzl-09, dipiazza-12, heinzl-12}.
The existence of background fields lead to surprising effects: 
the most known example is the Schwinger mechanism \cite{sauter-31, heisenberg-36, schwinger-51}, where electron-positron pairs are spontaneously produced from the quantum vacuum fluctuations stimulated by the background field.

Theoretical studies of interactions in external background fields are mostly considered in the quantum mechanical Furry picture \cite{furry},
which provides a modified representation of the usual Dirac picture.
The modification in this approach is caused by adding the external field to the free Hamiltonian.
In case of the Dirac equation, for instance, if the background is supposed to be a plane-wave, these dressed state solutions are known as the so-called Volkov representations \cite{volkov}.  
More on analogous semi-classical techniques will be discussed in Sec.~\ref{sec_semiclass} below. 
On the contrary, in Refs.~\cite{bergou-80, bergou-81a, bergou-81b}, the authors have shown for the photon scattering process, see Sec.~\ref{sec_scattering-gen}, that
considering a radiation field as fully quantised leads under certain conditions to the same result obtained with the semi-classical approach.

After the invention of novel laser technologies much progress was made for strong-field pair production \cite{reiss} and for strong-field photon scattering \cite{nikishov-64, nikishov-65, goldman-64, narozhnyi-65, brown-64, kibble-65}, particularly in specific plane-wave fields. 
Many other processes have mainly been studied either in mono-frequent or constant plane-wave fields since backgrounds with such a space-time dependence allow mathematically an exact treatment of the interaction, also because those field geometries can suitably be applied for modelling electromagnetic fields of laser beams \cite{bamber}.

Further strong-field phenomena have been already comprehensively investigated: for instance the laser-assisted pair production process in the field of a nucleus \cite{mittleman-87, muller-03, milstein-06, kuchiev-07, kaminski-06}. Other second-order processes which are related to two-loop self energies have been considered, including laser-assisted M\o ller scattering \cite{oleinik-67, roshupkin-96, panek-04}, laser-assisted Mott scattering \cite{szymanowski-97, panek-02}, laser-assisted Bremsstrahlung \cite{loet-07, schnez-07}, laser-assisted Compton scattering \cite{oleinik-68, belousov-77} and double-photon Compton 
scattering \cite{morozov-75, loet-09b}. Furthermore, effects on the weak interaction in the presence of strong laser fields have been studied for neutrino production \cite{titov-11} and the muon decay \cite{dicus-09, farzinnia-09}. For a further overview on processes in SFQED we refer to the reviews in Refs.~\cite{ritus, mcdonald-86, salamin-06, marklund-06, ehlotzky-09, dipiazza-12} an the references therein.

Describing the external background as simplified plane-waves is a common approach which can be enlightening for studying interesting properties of SFQED, however, other essential features might be still undetected. The importance of considering different field configurations has been recently stressed in Refs.~\cite{schuetzhold-08, schneider, linder, orthaber-11, otto-15, otto-15-v2, hebenstreit-09, dumlu-10, akal-14} for Schwinger pair production, for the assisted nonlinear Breit-Wheeler process \cite{jansen-13} and for radiation reaction and vacuum birefringence \cite{torgrimsson-16}. Studying a variety of field shapes might be therefore useful to increase the precision of theoretical predictions as well as to optimise the background geometry for upcoming experimental setups \cite{kohlfuerst-13}.

From the experimental point of view, fundamental SFQED processes have been considered for the first time in the E144-experiment at SLAC \cite{burke-97, bamber} where electrons with energies around $46.6$~GeV collided with a laser beam of intensity about $10^{18}$~W/cm$^2$. The following two processes were observed taking place in the external laser field:
photon scattering by an electron (nonlinear Compton scattering), cf. Sec. \ref{sec_scattering-gen}, and stimulated electron-positron pair
production, cf. Sec. \ref{sec_pp}. For recent studies of these processes we refer to Refs.~\cite{heinzl-09, harvey-09, heinzl-10, mackenroth-10, seipt-13, dinu-13, seipt-16} and \cite{heinzl-marklund-10, titov-12, nousch-12, jansen-15}, respectively. 
Additional polarisation effects have been investigated in Refs.~\cite{ivanov-cs, ivanov-pp}. 
For studying processes taking place in such external backgrounds, it is mostly common to characterise the laser field by an invariant classical intensity parameter $\xi$ which will be discussed in more detail in Sec.~\ref{subsec_genSmatrix} below. In SLAC-E144 \cite{burke-97, bamber} a relatively low intensity parameter, e.g. $\xi < 1$, had been applied, whereas present modern lasers can have intensity parameters $1 \leqslant \xi \leqslant 10$ and beyond. Electrons in such laser environments become already highly relativistic during one laser period. Because of this, large intensity parameters, i.e. $\xi > 1$, forbid to treat the background field perturbatively.

So far, further predicted strong-field phenomena as light-by-light scattering or 
vacuum birefringence~\cite{heinzl-06, karbstein} could not be tested. However, the currently planned and built modern strong-field laser facilities as, for instance, CLF \cite{clf}, ELI \cite{eli}, ELI-NP \cite{eli-np}, XCELS \cite{xcels} and the European XFEL \cite{xfel} are promising in this regard and new exciting times are just ahead. A more detailed overview can be found also in \cite{dipiazza-12}. 
In addition to those promising constructions, referring in particular to the ALPS \cite{alps} experiment, even further new 
surprising phenomena like minicharged particles or the effects of the 
axion-particle might be accessible \cite{jaeckel-10, bahre-13, gies-09, chavez-14}. 
Meanwhile, it has been also pointed out that strong electromagnetic field effects can become important in the interaction region of future linear colliders, e.g. ILC and CLIC \cite{hartin-12, porto-13, hartin-14}.
Strong fields can be also used to test aspects of theories with extended space-time structure, e.g. space-time noncommutativity \cite{heinzl-ilderton-10}. Hence, all in all, those promising future facilities will give the opportunity to test a variety of elementary features as well as important non-perturbative predictions in quantum field theory.

In order to study more general properties of processes in SFQED, it
would be therefore useful to have analytic expressions valid for a whole class of
generalised external background fields.
The advantage of
such general analytic expressions is that not only well established results
for particular external fields 
can be immediately obtained as special cases,
but also the total probability
for external fields not yet studied
would be theoretically available. Apart from considering different field geometries, it would be also beneficial 
to investigate interference effects due to phase variations in the field modes and possible enhancement signatures, as, for instance, realised through dynamical assistance in different pair production scenarios. 
Taking those aspects into account, we focus in this paper on elementary nonlinear phenomena and
derive expressions for total probabilities in case of nonlinear Compton scattering and stimulated lepton-antilepton pair production --- as for instance already considered in Ref.~\cite{bamber} --- for a class of generalised
external electromagnetic fields and we illustrate the formulae for particular special cases.

The paper\footnote{Note that units with $\hbar = c = 1$ and Minkowskian space-time with signature $\mathbf{R}^{1,3}$ are employed throughout this paper.} is structured as follows:
In Sec.~\ref{sec_semiclass} we briefly address a semi-classical technique based on coherent states of radiation. According to this method the background can principally be treated as a classical field.
In Sec.~\ref{sec_dirac-fermions} we present the basic properties of Dirac fermions in electromagnetic plane-wave fields and discuss important characteristics particularly in the presence of fields with periodic space-time dependence.
In Sec.~\ref{sec_scattering-gen} we derive a closed analytical expression for the total Compton scattering probability for a whole class of generalised background fields.
In Sec.~\ref{sec_compton_nc} we apply our general Compton scattering formula 
containing additional terms in comparison to usual expressions
to the case of propagating electron in a specific laser-like background 
without imposing any constraints on the field photon energies
and discuss the relevance of those extra contributions.
In Sec.~\ref{sec_pp} we adopt our result to the process of stimulated lepton-antilepton pair production in photon collisions. We derive again a closed formula for the total pair production probability without restricting interacting photon energies.
Sec.~\ref{sec_out} gives a short summary of our studies.
All relevant abbreviations and the integral solutions are attached in the Appendices.
Furthermore, we explicitly show how to obtain the well-known soft-photon-limit (SPL) for Compton scattering, starting from our general formula. 
\section{Semi-classical approach to QED on external backgrounds}
\label{sec_semiclass}
We consider interactions taking place in strong external electromagnetic background fields with a generalised form. 
Principally, it is difficult to define how many photon quanta of the external field have interacted. Therefore, a complete quantum field theoretical approach has not been performed in the literature. Nevertheless, due to the expected huge
photon number in a strong field appearing for instance in modern laser facilities, it should however be feasible
to treat the field classically, as some fixed background.
Describing the strong bosonic background in terms of 
coherent states $|C\rangle$ of radiation, also called ``Glauber states'' \cite{glauber1,glauber2,glauber3}, will allow 
to proceed in that way. 
Those states are based on the
expectation value of the photon number operator,
a photon creation operator $\hat{a}^\dagger_\mu$ and a polarisation and momentum distribution function of
the field photons $C_\mu$ \cite{schwinger,glauber1,glauber2,glauber3,klauder-85,combescure-12}.

Note that classical fields and such coherent states are naturally related, where latter one are taken as the ``most classical'' wave functions satisfying the minimal uncertainty relation such as the quantum mechanical ground state $|0\rangle$.
The corresponding classical field is nothing but the Fourier transformation of the coherent distribution function \cite{harvey-09}. 
We will underline this in the following by considering S-matrix elements.

S-matrix elements between states including coherent states
are equivalent to a particular weighted sum over S-matrix elements of associated photon states in Fock space. 
Furthermore, these kind of states can be seen physically as neglecting depletion of the background field. 
This allows immediately to consider the number of field photons as constant \cite{birula,bergou-81b}.
Let us suppose an arbitrary scattering process with an asymptotic in-state containing a coherent state $|C\rangle$ and some
other undefined particle states. Assuming all of those particles being part of the
coherent state will lead to technical difficulties, see Ref.~\cite{harvey-09}. Therefore, the initial state shall be separated as $| \text{in}; C \rangle$, whereas
the out-state is supposed to be of the form $\langle \text{out}; C |$ since no photon depletion is expected. In other words, one has to determine 
matrix elements $\langle \text{out}; C |  \hat{\mathcal{S}} | \text{in}; C \rangle$ with respect to the following S-matrix operator,
\eqn{
\hat{\mathcal{S}} \equiv \mathcal{T} \exp \lc -i  \int_{-\infty}^\infty dt \hat{H}_I (t) \rc,
}
whereas $\hat{H}_I (t)$ is the interaction Hamiltonian and $\mathcal{T}$ denotes the usual time-ordering operator. By setting $|C\rangle$ as a translation of the vacuum state $|\text{VAC}\rangle$ \cite{gottfried} --- i.e. $|C\rangle = \hat{T} |\text{VAC}\rangle$, therefore the coherent state is also called ``displaced ground state'' --- we extract\footnote{$C_\mu(\pmb{k}^\prime) = 0$ since no forward scattering assumed} the translation operator from the states.
This leads to an ordinary asymptotic Fock state and a
modified S-matrix operator as follows
$\langle \text{out}; C |  \hat{\mathcal{S}} | \text{in}; C \rangle = \langle \text{out} | \hat{T}^{-1}  \hat{\mathcal{S}} \hat{T} | \text{in} \rangle$ \cite{harvey-09}.
According to $\hat{\mathcal{S}}$ the translation operators shift any photon operator $\hat{A}_\mu$ appearing in the interaction Hamiltonian by $C_\mu$ due to a Fourier decomposition denoted by $\gA_\mu$. This has the consequence that leptons interact with the full quantised photon field $\hat{A}_\mu$ plus a classical background field $\gA_\mu$.
As an immediate consequence, associated Feynman diagrams will be
constructed by the quasi-classically modified action
\eqn{
S = \int d^4 x - \frac{1 }{4} F_{\mu \nu} F^{\mu \nu} + \overline{\Psi} \lL    i\slashed{\partial} - e \slashed{\gA} - e \slashed{\hat{A}}   - m   \rL  \Psi,
}
where the photon field in the interaction term of the ordinary QED action is now shifted by $\gA_\mu$.
Due to the reason that only the interaction terms are directly influenced by the background field,
all the information about asymptotic photon distributions will be taken into account
by inserting simply a classical field in the action. Accordingly, we may write
$\langle \text{out} | \hat{T}^{-1}  \hat{\mathcal{S}} \hat{T} | \text{in} \rangle \equiv \langle \text{out} |  \hat{\mathcal{S}}_{\gA}  | \text{in} \rangle$,
where in- and out-states are meant to be ordinary particle number states and
the quantised photon field in $\hat{\mathcal{S}}$ is shifted by the classical background field $\gA_\mu$ \cite{kibble-65}.

\begin{figure}[h]
\centering
\includegraphics[width=0.9\textwidth]{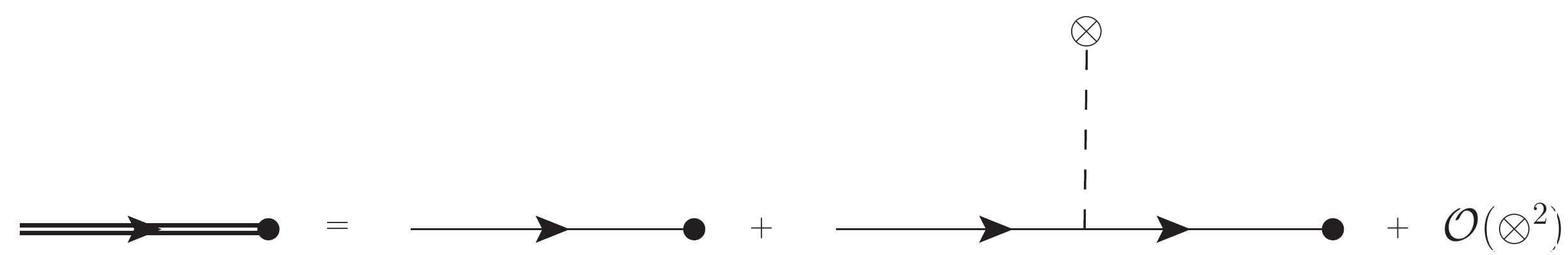}
\caption{Perturbative decomposition of a ``dressed'' fermionic external line for the lepton in external background field. The expansion is in terms of a free external line interacting an infinite number of times with the background field, shown by the dashed line connection to $\otimes$.}
\label{fig_lep-prop}
\end{figure}

Latter action is obviously quadratic in the lepton fields such that all background induced effects are
included in a ``dressed'' fermionic line by $\gA^\mu$ which is
in diagrammatic description mostly represented
by a double line. Due to reasons which will become clear below, the field dressing is shown schematically in Fig.~\ref{fig_lep-prop} on the basis of a dressed external line. 
It is expanded in terms of free lepton external lines interacting an
infinite number of times with $\gA^\mu$ illustrated by some dashed lines.
It is remarkable that Feynman rules are the same as in ordinary QED.
Particularly, at leading order, in case of electron-photon scattering discussed below, we consider only a single three vertex connection joining one photon external line and two of the
``dressed'' fermionic external lines for the in- and out-going electron at tree-level. 
Therefore, we do not need to consider any internal lines (propagators) in forthcoming calculations. Nevertheless, it is worth mentioning that internal lines are analogously replaced by some ``dressed'' lines, where similarly the free fermionic propagator, for instance in case of a periodic plane-wave background, is sandwiched by some external field induced contributions \cite{porto-13}. However, this leads to non-trivial dependencies on space-time coordinates such that space-time locality is not anymore conserved. Anyway, in present paper we will not address those conceptual difficulties.

As a side note following consideration might provide some alternative insights into present described approach.
Namely, the fact of
translated photon fields in the S-matrix operator by the classical background field
emphasizes rather that the external field will influence somehow the way how particles interact with the environment, say their interaction strengths with the field bosons.
The corresponding modification on interaction level will be therefore encoded in those dressed lines.
In this context, we will be pointing out later on, that this dressing might be more affecting the background exactly in the interaction points, i.e. the usual QED vertices. Principally, those could be interpreted as dressed whereas the connected external lines remain as the usual free one. This will be naively in good agreement with the mentioned violation of locality during for instance a two vertex process including a fermionic internal line which depends non-trivially on both space-time points $(x,y)$ separately instead on their difference $(x-y)$. 

Furthermore, according to the mechanism of dynamical assistance in pair production, this kind of interpretation becomes more reasonable. Namely, in case of Schwinger pair production, the production rate can be substantially enhanced by superimposing one weak rapidly varying mode with a strong slowly varying one. The energy threshold will be narrowed by the weak mode such that pair production can be stimulated by the remaining mode \cite{schuetzhold-08}. This illustrates as well the idea of a shifted ``vacuum'' frame caused by the external field.   

One should remind that it is in general very difficult to construct a closed expression for the lepton propagator. Since an intense background will be
identified by large values of characteristic quantities\footnote{for instance the classical field intensity parameter $\xi$ --- see Sec.~\ref{subsec_genSmatrix}\label{fnote:xi}}, a perturbative treatment of the
background is not feasible, even not if one performs
some coupling expansion. Fortunately, for the class of backgrounds with plane-wave character considered
in this paper, the dressed lepton states, say for the electron, are known exactly in form of Volkov representations \cite{volkov}. This allows consequently to treat the background
field exactly. 

Moreover, in recent studies, exact analytical state solutions
have been constructed under certain energy scale assumptions even for backgrounds with general space-time structure \cite{diPiazza-14}.
However, we will continue illustrating latter ideas by applying them to nonlinear lepton-photon interaction processes keeping the plane-wave background field shape general.

\section{Dirac fermions in electromagnetic plane-wave fields}
\label{sec_dirac-fermions}

\subsection{General remarks and exact solutions}

This section reminds some basic aspects of describing particle states in arbitrary external electromagnetic plane-wave fields and introduces 
the exact Volkov solutions \cite{volkov} of the Dirac equation in such external backgrounds.

In the following
let us consider an electromagnetic field $\gA$ characterised via the the four-component 
wave-vector $k^\mu$. 
Due to its plane-wave character
the field depends only on $\eta \equiv k \cdot x$\footnote{Here, we will use the Einstein summation convention $k \cdot x \equiv k^\mu x_\mu \equiv \sum_{\mu = 0}^{4} k^\mu x_\mu$ (note $x^2 \equiv x \cdot x$) for four-vectors and the short Feynman slash notation $\slashed k = \gamma^\mu k_\mu$ by making usage of the Dirac matrices $\gamma^\mu$.}, i.e. $\gA^\mu \equiv \gA^\mu (\eta)$.
We impose that the field satisfies the Lorenz gauge condition
$\partial_\mu \mathcal{A}^\mu(\eta) = k_\mu \partial_\eta[\mathcal{A}^\mu(\eta) ] = \partial_\eta[k_\mu \mathcal{A}^\mu(\eta) ] = 0$,
where we can set without loss of generality $k_\mu \mathcal{A}^\mu(\eta) = 0$, since
$k_\mu \mathcal{A}^\mu = \mathfrak{C} \in \mathbb{R}.$ 
The corresponding gauge bosons are assumed to be massless photons with $k$ being light-like, i.e. $k^2 = 0$.
The  field tensor is given by 
$F_{\mu \nu} = \partial_\mu \gA_\nu(\eta) - \partial_\nu \gA_\mu(\eta) = k_\mu \partial_\eta [\gA_\nu(\eta)] - k_\nu \partial_\eta [\gA_\mu(\eta)]$.
The second order Dirac equation, obtained by applying $\gamma^\mu[\hat p_\mu - e\gA_\mu] + m$, in the presence of the external field becomes, cf. Ref.~\cite{landau},
\begin{equation}
\lL -\partial^2  - 2 i e \gA(\eta) \cdot \partial + e^2 \gA^2(\eta) - m^2 - ie \slashed k \cdot \slashed \gA(\eta)  \rL \Psi(\eta) = 0,
\label{eq_dirac}
\end{equation}
where $m$ is indicating the rest mass of the lepton (electron), $e$ denoting its electric charge and $\partial^2 = \partial_\mu \partial^\mu$.  A solution of this equation can be derived
in the form 
\eqn{
\Psi(\eta) = \exp(-i p \cdot x) F(\eta),
}
where $p^\mu$ denotes a constant four-vector.
It is obvious
that the function $\Psi$ will be unaltered if adding  any constant multiple of $k^\mu$ to $p^\mu$
and refining the function $F(\eta)$ appropriately. Therefore, one can additionally impose without loss of generality the particle momentum being on-shell, e.g. $p^2 = m^2$.
Solving the Dirac equation one obtains the general Volkov solution \cite{volkov}
\begin{equation}
\Psi^V_{p} =  \Lambda(\gA|\eta,p)  u_p  \exp \lc -i \mathcal{F}_p(\eta) - i p \cdot x \rc,
\label{eq_defvol_p}
\end{equation}
where
 $u_p \equiv u(p)$ denotes the complex free bispinor in momentum space, $p_0$ the particle energy and $\Lambda(\gA|\eta,p)  \equiv \lL 1 + \frac{e \slashed k \cdot \slashed \gA(\eta)}{2 k \cdot p} \rL [2 p_0]^{-1/2}$ a field-dependent combination of Dirac matrices. The function $\mathcal{F}_p$ is given in its integral form as follows
$\mathcal{F}_p(\eta) \equiv \int_{0}^{\eta} \lL \frac{e p \cdot \gA(z)}{k \cdot p} - \frac{e^2 \gA^2(z)}{2 k \cdot p} \rL dz$ \cite{landau}.
Concerning the bispinor $u_p$, it is supposed that the external field is switched on with infinite slowness from $t = - \infty$ 
such that due to $\gA(\eta) \rightarrow 0$ for $\eta \rightarrow -\infty$ the wave function $\Psi_p^V$ is identical to the solution $\Psi_0$ for the free Dirac's equation. Therefore $[ \slashed p - m ] u_p = 0$ is fulfilled.
Since the bispinor $u_p$ is time-independent, this statement holds even for a finite $\eta$. Thus, the resulting bispinor amplitude will be the same as the free one such that we shall take it to be normalised by the same condition $\overline{u_p} u_p = 2m$
, where $\overline{u_p} \equiv u_p^\dag \gamma^0$ is the usual Dirac adjoint.
The infinitely slow application of the field does not alter the normalisation integral and therefore the same condition
as for the free field solution will be imposed: 
$\int d^3 x\  \overline{\Psi_{p_f}^V} \gamma^0\Psi_{p_i}^V = (2\pi)^3 \delta^{(3)}(\pmb{p}_f - \pmb{p}_i)$.
%

\subsection{Particle interaction in periodic background fields}

The current density corresponding to the wave
function from Eq.~\eqref{eq_defvol_p} is given by
\begin{equation}
j^\mu = \overline{\Psi_p^V} \gamma^\mu \Psi_p^V =
  \frac{1}{p_0} \lL p^\mu -e\gA^\mu(\eta) + k^\mu \lL \frac{e p
    \cdot \gA(\eta)}{k \cdot p} - \frac{e^2 \gA^2(\eta)}{2 k \cdot p}
  \rL \rL,
\label{eq_current}
\end{equation}
where $k^2 = \slashed
k^2 = 0$, $\slashed \gA^2(\eta) = \gA^2(\eta)$ and $k \cdot \gA(\eta)
= 0$ has been used, cf. Ref.~\cite{landau}.
For periodic functions the time-averaged values are zero, i.e. $\langle \gA(\eta)  \rangle \equiv 0$,
such that the mean value of the current density can be obtained as
$\langle j^\mu \rangle = \frac{1}{p_0} \lL p^\mu  - \frac{e^2 \langle \gA^2(\eta) \rangle}{2 k \cdot p} k^\mu  \rL$.
In a next step the kinetic momentum density is calculated 
$\overline{\Psi_p^V}\gamma^0(\hat p^\mu - e \gA^\mu(\eta))\Psi_p^V$, applying
the momentum operator $\hat p^\mu - e \gA^\mu(\eta) = i \partial^\mu -
e \gA^\mu(\eta)$ explicitly, and one obtains
for the time-averaged value of the effective momentum
\eqn{
q^\mu = p^\mu - \frac{e^2 \langle \gA^2(\eta) \rangle}{2 k \cdot p} k^\mu.
\label{eq_defq}
}
The squared value yields $q^2 = m_{\star}^2$
with $m_{\star} = m \sqrt{1 - \frac{e^2 \langle \gA^2(\eta) \rangle}{m^2}}$.
The shifted quantity
$m_{\star}$ could be interpreted therefore as an effective quasi-mass of the particle
in the external background.
Note that this quasi-mass only applies, strictly speaking, in the interaction point, as pointed out later. It is rather a mathematical consequence of the external background periodicity which regulates somehow the strength of the interaction. This fact is more or less indicated by the Volkov solution itself. External field induced quasi-momenta are namely acting as a phase in the exponential whereas the Dirac bispinor remains unaffected. Normalising the states as previously discussed can thus be seen as an adapting procedure to the modified background.
The resulting
particle interaction is therefore influenced and modified by the field
shape of the present background. 
Applying Eq.~\eqref{eq_defq}
one obtains $\langle j^\mu \rangle = q^\mu / p_0$.
Due to this simple rescaling, the previously
presented normalisation condition becomes on basis of the $q$-scale
\eqn{ \int d^3 x\ \overline{\Psi_{p_f}^V}\gamma^0
  \Psi_{p_i}^V = (2\pi)^3 \delta^{(3)}(\pmb{q}_f - \pmb{q}_i)
  [q_0/p_0],
} 
where $q_0$ is being the effective electron energy influenced by
the external field. 
One has to note that according to Eq.~\eqref{eq_defq} especially the shape of $\langle \gA^2 \rangle$ plays a rather important role.
In cases, where $\langle \gA^2 \rangle$ is a time-independent fixed constant,
the transition from the classical momentum $p^{\mu}$ to the
quasi-momentum $q^{\mu}$ can be considered as a simple translation caused by some constant times $k^\mu$. 
Consequently, the only change for passing into a specific momentum scale will be the
normalisation of the wave function on the corresponding scale which is
simply characterised by choosing $p_0$ or $q_0$, respectively, in the
Volkov solution, cf. also Ref.~\cite{greiner}.
For being consistent with previously performed calculations in case of some very specific field 
configurations in Refs.~\cite{landau,ritus}, we will choose in the following the wave function on the $q$-scale.
Moreover, it is worth mentioning that by considering a scattering process, for instance, a simple two-particle
process, like pair production out of the Dirac vacuum by photon-photon collision,
the change between the two different scales would lead 
to the factor $d^3 \pmb{q} = [q_0 / p_0]\ d^3 \pmb{p}$
in the corresponding phase-space, cf. Ref.~\cite{greiner} and also Sec.~\ref{sec_pp} for more detail.
%
\section{Scattering in generalised plane-wave background fields}
\label{sec_scattering-gen}
%

\subsection{General S-matrix element}
\label{subsec_genSmatrix}

In the following part we will reduce our considerations to periodic plane-wave backgrounds of the form
\eqn{
\gA^\mu(\eta) = \sum_{r=1}^N a_1^\mu {A_1}_r(\eta) + a_2^\mu {A_2}_r(\eta),
\label{eq_gen-field}
}
with general $2\pi L$-periodic functions ${A_1}_r$ and ${A_2}_r$ in $\eta$, whereas $L \in \mathbb{Q}_+$ and $N \in \mathbb{N}$ shall be kept arbitrarily.
Furthermore, according to previous discussion on the role of $\langle \gA^2 \rangle$, we impose the condition 
\eqn{
\int_0^\eta \gA^2(z)\ dz \cong \eta \langle \gA^2(z) \rangle = \eta  \mathfrak{a}^2 
\label{eq_A2-constraint}
}
in Minkowskian space-time, where $\mathfrak{a}^2 < 0$ is the previously mentioned time-independent constant.
Introducing a Lorentz-invariant dimensionless field intensity parameter of the form  
\eqn{
\xi^2 \equiv \frac{- e^2 \mathfrak{a}^2}{m^2},
\label{eq_dim-param}
}
the corresponding effective quasi-momentum of the electron, embedded in the external background, is given by
\eqn{
q^\mu = p^\mu + \frac{m^2 \xi^2}{2 k \cdot p} k^\mu
\label{eq_defq_2}
} 
and the shifted effective mass-like quantity for the electron can be defined as
\eqn{
m_{\star} = m \sqrt{1 + \xi^2}.
\label{eq_effm_2}
}
The parameter $\xi$ is a well known characteristic parameter often used in the literature: it comprises the so-called multi-photon processes
and provides a measure for the intensity of the field and is therefore 
suitable for distinguishing the different interaction regimes. 
However, note that in general a parameter, defined as $\frac{-e^2 \langle \mathcal{A}_\mu \mathcal{A}^\mu  \rangle}{m^2}$, establishes Lorentz invariance but does not conserve gauge invariance in any arbitrary gauge. Fortunately, for the plane-wave approach considered in this paper, the previously defined $\xi$ in \eqref{eq_dim-param} is a fully invariant parameter. For more on this issue we refer the reader to Ref.~\cite{heinzl-08}.
The leading order term for the resulting S-matrix element $\mathcal{M}_{fi}$ for the scattering process
of a photon by an electron 
in the external background field is given as follows:
\eqn{
\mathcal{M}_{fi} = \frac{-ie}{\sqrt{2 \omega_f}} \int d^4x\ \exp(i k_f \cdot x) \overline{\Psi^V_{ q_f }} \slashed{\varepsilon}_f \Psi^V_{ q_i } ,
\label{eq_mif}
}
where $\varepsilon_f^\mu$ denotes the polarisation vector of the final emitted photon. Due to reasons which will become clear shortly, a diagrammatic illustration of $\mathcal{M}_{fi}$ will be presented later in Fig.~\ref{fig_compton} using previously introduced Volkov external lines from Fig.~\ref{fig_lep-prop}. 
Note that the exact Volkov solution, normalised with respect to the $q$-scale, simply reads as
\eqn{
\Psi^V_{ q } 
=  \Lambda(\gA|\eta,q) u_p  \exp \lc -i \mathcal{P}_q(\eta) - i q \cdot x \rc,
\label{eq_defvol_q}
}
where $u_p$ is still the free Dirac bispinor and the function $ \mathcal{P}_q(\eta) \equiv \int_0^{\eta} e q \cdot \gA(z) [k \cdot q]^{-1} dz$ has been introduced by replacing the function $\mathcal{F}_p (\eta)$ with the help of our assumption\footnote{\label{fnote_A2cond}This ensues that we can apply the present formula for a large class of fields, including also the case of linear polarisation, which fulfil the condition \eqref{eq_A2-constraint} as well.} from Eq.~\eqref{eq_A2-constraint} and the Lorenz gauge correspondingly. 

\begin{figure}[h]
\centering
\includegraphics[width=0.4\textwidth]{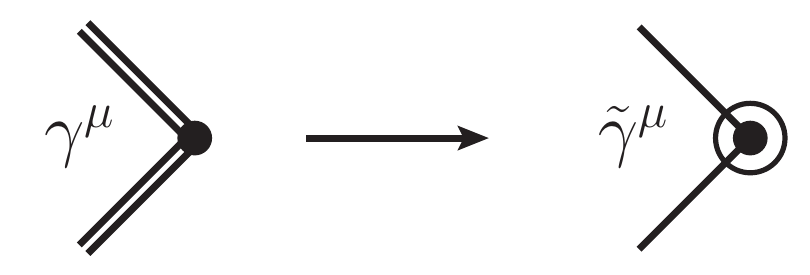}
\caption{Transition from free vertex $\gamma^\mu$ sandwiched between two Volkov lines to a ``dressed'' vertex $\tilde{\gamma}^\mu$ which is basically connecting two free fermionic lines instead.}
\label{fig_line-vertex}
\end{figure}

Applying Eq.~\eqref{eq_defvol_q} for the initial electron state $\Psi^V_{ q_i }$ and for the final state $\overline{\Psi^V_{ q_f }}$ in its Dirac adjoint we derive 
the S-matrix element $\mathcal{M}_{fi}$ according to Eq.~\eqref{eq_mif} as follows:
\eqnsplit{
&\mathcal{M}_{fi} = -ie \frac{1}{\sqrt{2 \omega_f}}  \frac{1}{\sqrt{2 q_{i,0}}}  \frac{1}{\sqrt{2 q_{f,0}}} \int d^4 x\ \exp\lc -i(q_i - q_f - k_f) \cdot x \rc\\
&\times  \overline u_{p_f}  \exp \lc i \mathcal{P}_{q_f}(\eta) \rc  
\lL 1 + \frac{e  \slashed \gA(\eta) \cdot \slashed k }{2 k \cdot q_f } \rL
\gamma^\mu  
\lL 1 + \frac{e \slashed k \cdot \slashed \gA(\eta) }{2 k \cdot  q_i } \rL
\exp \lc -i \mathcal{P}_{q_i}(\eta)  \rc  
{\varepsilon_f}_\mu  u_{p_i}.
\label{eq_matrix-cs}
}
Note that the original free vertex $\gamma^\mu$ in pure QED vacuum sandwiched by two Volkov external lines could be principally interpreted as a modified vertex by the external background field, shown in Fig.~\ref{fig_line-vertex}, e.g. cf. \cite{mitter-75, meuren-13, meuren-15}.
That resembles a transition $|\text{VAC}\rangle_{\text{QED}} \longrightarrow |\text{VAC}\rangle_{\text{EXT}}$ from the QED vacuum to a background affected shifted ``vacuum'' frame in the interaction point.
This kind of reinterpretation seems to be more suitable for an intuitive understanding according to scattering processes with asymptotic free states as usually supposed. In this way, one may accept the resulting field, encoded in the Volkov solutions, as a modified background environment exactly in the interaction point but not as some effective dressing during the propagation of particles in the external field. The particles could still be understood as the usual free states. Therefore, it might be more appropriate to 
interpret the effects of the external field as a ``dressed vacuum'' instead
of dressing the particles with an effective mass-shift.
The corresponding adapted ``dressed'' vertex function reads as:
\eqn{
\tilde{\gamma}^\mu( q_f , q_i, \eta) 
= \exp \lc i \mathcal{P}_{q_f}(\eta) \rc    \Pi^\mu_{\eta,cs} ( q_f , q_i )    \exp \lc - i  \mathcal{P}_{q_i}(\eta)  \rc,
\label{eq_vert_dres}
}
where
\eqn{
\Pi^\mu_{\eta,cs} ( q_f , q_i ) \equiv \lL 1 + \frac{e  \slashed \gA(\eta) \cdot \slashed k }{2 k \cdot q_f } \rL  \gamma^\mu  \lL 1 + \frac{e \slashed k \cdot \slashed  \gA(\eta) }{2 k \cdot q_i } \rL
\label{eq_pi}
}
has been separated 
for a better readability and some simplifications (subscript $cs$ stands for Compton scattering).
Applying a usual Fourier expansion\footnote{\label{fnote_special-characters}Note that from now on special characters like $\varphi$ or $\vartheta$ will be used for the Fourier phases whereas usual symbols $\phi$ and $\theta$ denotes the azimuth and polar phase-space angle, respectively.} on the $2 \pi L$-periodic vertex function  $\tilde \gamma^\mu ( q_f, q_i ,\eta)$, one obtains finally
\eqnsplit{
\tilde\gamma^\mu (q_f , q_i ,\eta) 
&= \frac{1}{2 \pi L}  \sum_{n = - \infty}^{\infty}  \int_{- \pi L}^{ \pi L}   \exp \lc i s_n (\varphi - \eta)  \rc  \tilde\gamma^\mu ( q_f ,q_i ,\varphi)\ d\varphi
\label{eq_vert-four}
}
with the abbreviation $s_n \equiv \frac{n}{L}$.

\begin{figure}[h]
\centering
\includegraphics[width=0.35\textwidth]{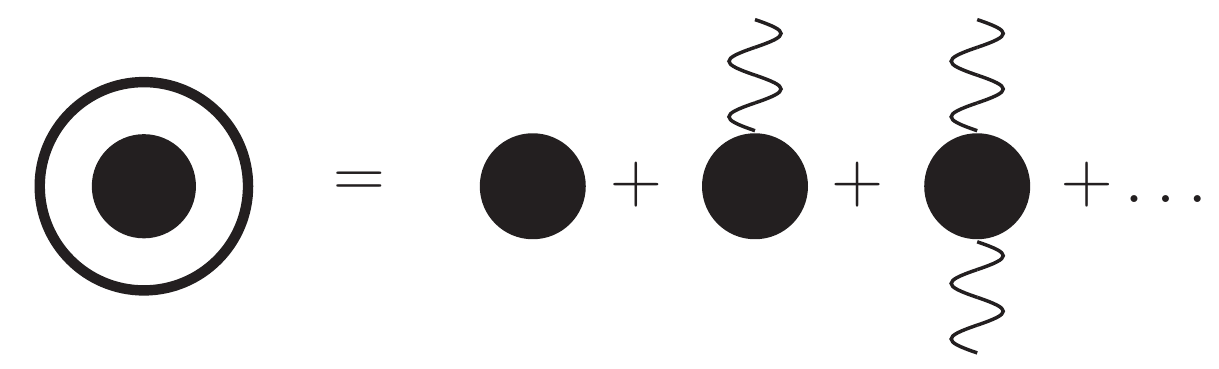}
\caption{Due to Fourier decomposition the dressed vertex $\tilde\gamma^\mu$ could naively be seen as the usual free one surrounded by external field photons with wave vector $k^\mu$ of infinite number, represented by additional wavy lines.}
\label{fig_vertex}
\end{figure}

Such a spectral decomposition of the ``perturbed'' vertex will reduce difficulties in the calculation. Therefore, although we have started with the exact non-perturbed Volkov solutions in the external field, 
the resulting expressions will nevertheless acquire a perturbative behaviour corresponding
the decomposition in Eq.~\eqref{eq_vert-four}. In particular, the interpretation of ``dressing'' the usual free vertex by external field photons with wave-vector $k^\mu$, Fig.~\ref{fig_vertex}, becomes more clarified.

Consequently, a full diagrammatic scheme of the scattering process at tree-level can be presented as in Fig.~\ref{fig_compton} (b). Note that this kind of diagram provides an alternative way to the usual illustration by Volkov external lines as shown in Fig.~\ref{fig_compton} (a).

\begin{figure}[h]
\centering
\begin{minipage}{0.45\linewidth}
\centering
\subfigure[``Dressed'' fermionic external lines and free vertex.]{\includegraphics[width=0.8\textwidth]{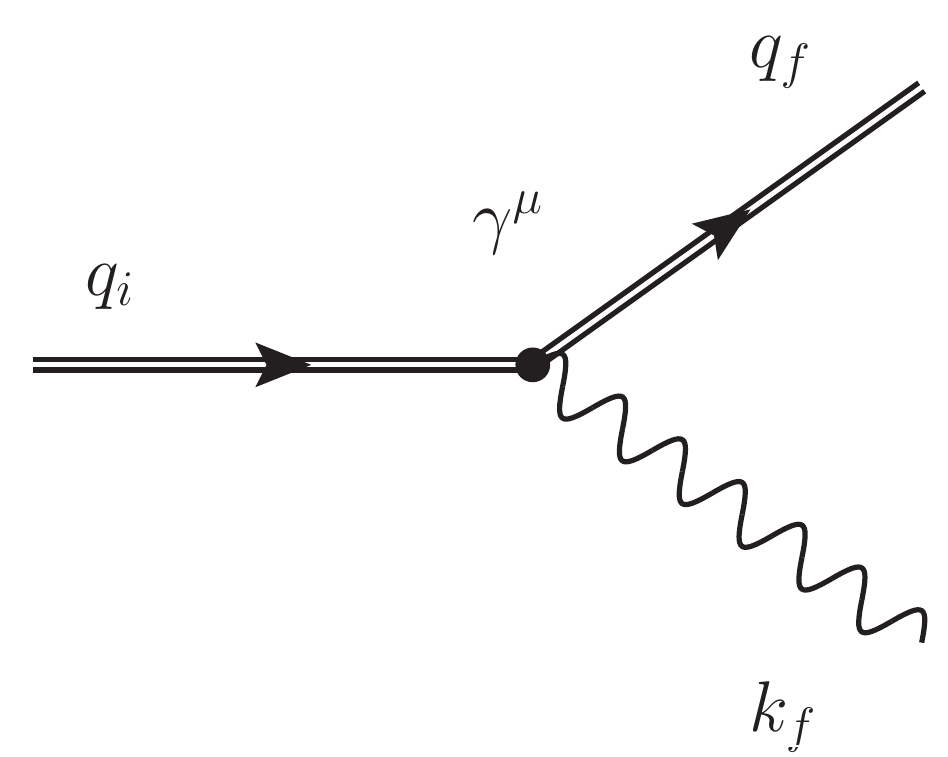}}
\label{subfig_dressed_vertex-a}
\end{minipage}
\begin{minipage}{0.45\linewidth}
\centering
\subfigure[Usual free fermionic external lines and ``dressed'' vertex.]{\includegraphics[width=0.8\textwidth]{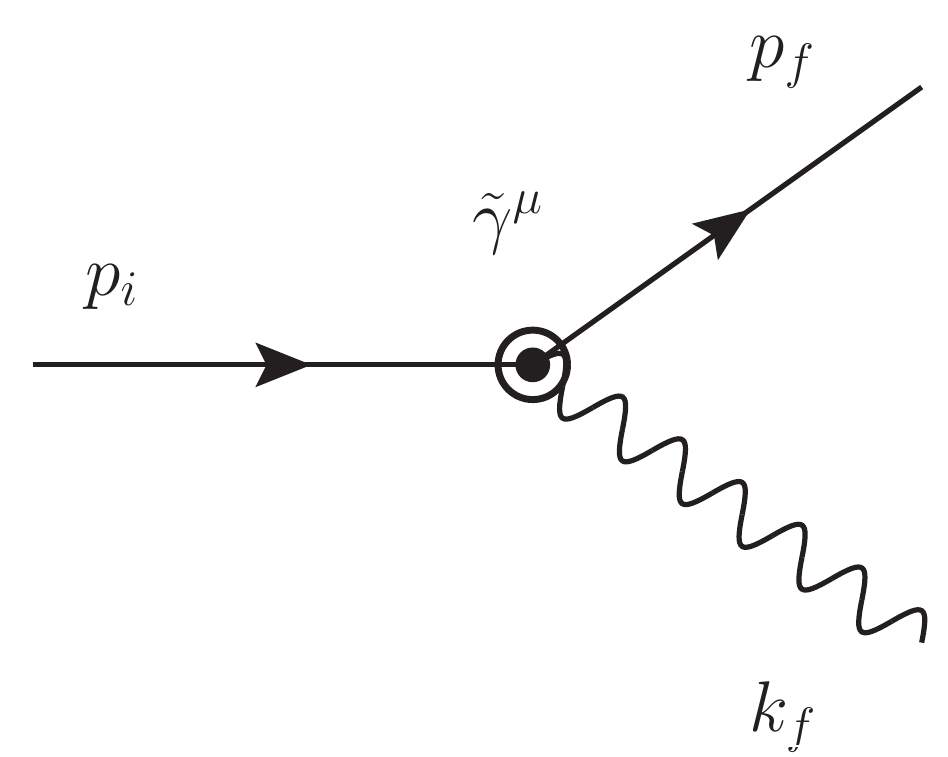}} 
\label{subfig_dressed_vertex-b}
\end{minipage}
\caption{Alternative diagrammatic illustrations for electron-photon interaction (Compton scattering) embedded in external background field. The modified ``dressed'' vertex $\tilde{\gamma}^\mu$ is denoted by an additional cycle as shown in Fig.~\ref{fig_vertex}. Double lines are indicating the exact fermionic Volkov external lines given for in-going and out-going electron states. The wavy line is representing the final emitted photon external line. Note that interpretation (b) becomes more convenient in the context of a ``dressed'' vertex function discussed above, but both fermionic external lines are as free states.}
\label{fig_compton}
\end{figure}

Based on Eq.~\eqref{eq_vert-four},
it has been directly received that the only position-dependent part in the integral \eqref{eq_matrix-cs} remains as a simple exponential function. This allows to perform the spatial integration which yields an identity of the following form
\eqn{
\int d^4 x\ \exp \lc -i \lL q_i + s_n k - q_f - k_f \rL \cdot x \rc = (2 \pi)^4 \delta^{(4)}\lc q_i + s_n k - q_f - k_f \rc,
\label{eq_delta-id}
}
where $\delta^{(4)}$ denotes the four-dimensional Dirac delta function. Inserting Eq.~\eqref{eq_delta-id}
into \eqref{eq_matrix-cs} leads finally to
\eqnsplit{
\mathcal{M}_{fi} &= -ie \frac{1}{\sqrt{ 2 \omega_f }}  \frac{1}{\sqrt{ 2 q_{i,0} }}  \frac{1}{\sqrt{ 2 q_{f,0} }} \frac{1}{2 \pi L} (2 \pi)^4  \sum_{n = - \infty}^{\infty}  \int_{- \pi L}^{ \pi L} d\varphi\\
&\times \delta^{(4)}\lc q_i + s_n k - q_f - k_f \rc
\exp \lc i s_n\varphi \rc
\overline u_{p_f}  \Pi^\mu_{\varphi,cs} ( q_f , q_i )  {\varepsilon_f}_\mu  u_{p_i}\\
&\times \exp \lc i [ \mathcal{P}_{q_f}(\varphi) - \mathcal{P}_{q_i}(\varphi) ] \rc.
\label{eq_Mfi-cs-sum}
}
Due to the summation in Eq.~\eqref{eq_Mfi-cs-sum} one could interpret $s_n$ as the net number of bosonic field quanta with momentum $k^\mu$ which are either absorbed ($n > 0$) or emitted ($n < 0$) in the process.
This result is somehow surprising, because originally the electromagnetic background field was introduced as a classical field and not as second-quantised.
The discretisation of photon momentum, evident in the Dirac delta function, just arises from the periodicity of the plane-wave background in space and time.
On the other hand the derived delta function forces a momentum conservation during the interaction
with respect to the number $s_n$ of field quanta, respectively.

\subsection{Differential scattering probability}

The differential probability for the scattering process per unit volume $V$ and unit time $T$ is simply given by a sixfold momentum integration of the form:
\eqn{
dW_{cs} = d^3\pmb{q}_f d^3\pmb{k}_f \sum_{\varepsilon_f,s_f} \frac{|\mathcal{M}_{fi}|^2}{VT},
\label{eq_dW-gen}
}
whereas the differential on the right hand-side for the final electron state is obviously due to the modified particle dynamics in the external field chosen on the $q$-scale. It should be noticed at this point that we aim basically calculating the probability for any spin state of the final electron and any polarisation of the emitted photon. Therefore, it is necessary to sum over all possible final spin $s_f$ and polarisation $\varepsilon_f$ configurations. Consequently, this requires calculating the square of the S-matrix element. 
Because of $|\mathcal{M}_{fi}|^2$, it is apparent that due to the multiplied delta functions the corresponding product $\sum_n \sum_r$ will force the equality $n = r$. Taking this into account, we apply the transition 
$(2 \pi)^8 \delta^{(4)}(\ldots)  \delta^{(4)}(\ldots) \longrightarrow (2 \pi)^8 \delta^{(4)}(\ldots)  \delta^{(4)}(0)$
and set $V T \equiv (2 \pi)^4 \delta^{(4)}(0)$, as usual.
Before continuing with computing the matrix square and thus $dW_{cs}$, we sum up over all spins and polarisation directions to keep those general. Let us first of all define two matrices $\mathrm{M}_{\varphi,cs} \equiv \Pi^\mu_{\varphi,cs} ( q_f , q_i)  {\varepsilon_f}_\mu$ and $\mathrm{M}_{\vartheta,cs} \equiv \Pi^\nu_{\vartheta,cs} ( q_f , q_i)  {\varepsilon_f}_\nu$. This yields the following summation
$\sum_{\varepsilon_f,s_f} \lL \overline u_{p_f}  \mathrm{M}_{\varphi,cs}  u_{p_i} \rL \lL  \overline u_{p_f}  \mathrm{M}_{\vartheta,cs}  u_{p_i}  \rL^*$
which leads to the usual trace, denoted by $\mathfrak{S}_{\mathrm{cs}}$, of the following form
$\text{Tr} \lc [\slashed p_f + m] \Pi_{\varphi,cs} [\slashed p_i + m] \gamma^0 {\Pi}^\dagger_{\vartheta,cs} \gamma^0 \rc$.
From the trace calculation one obtains the result
\eqnsplit{
\mathfrak{S}_{cs} &= 16 m^2 - 8 p_i \cdot p_f - 4 e \lL \gA_\varphi + \gA_\vartheta \rL \cdot \lL p_i \lL \frac{k \cdot p_f}{k \cdot p_i} - 1  \rL + p_f \lL \frac{k \cdot p_i}{k \cdot p_f} - 1  \rL  \rL\\
&- 4 e^2 \lL \gA^2_\varphi + \gA^2_\vartheta - \gA_\varphi \cdot \gA_\vartheta \lL  \frac{k \cdot p_i}{k \cdot p_f} + \frac{k \cdot p_f}{k \cdot p_i}  \rL \rL,
\label{eq_tr-result}
}
where the abbreviations $\gA_\varphi \equiv \gA (\varphi)$ and $\gA_\vartheta \equiv \gA (\vartheta)$ have been used. 
One should note that higher order terms in $e$ got cancelled in expression~\eqref{eq_tr-result} by applying the properties of $k^\mu$ and the Lorenz gauge condition.
Re-expressing $\mathfrak{S}_{cs}$ via the kinematically relevant momenta $q_i$, $q_f$, we obtain with 
\eqn{
p_i \cdot p_f = q_i \cdot q_f +  \frac{e^2 \mathfrak{a}^2}{2} \lL \frac{k \cdot q_i}{k \cdot q_f}  + \frac{k \cdot q_f}{k \cdot q_i} \rL
\label{eq_pTOq-cs}
}
and the Lorenz gauge for the differential probability \eqref{eq_dW-gen}:
%
\eqnsplit{
&dW_{cs} = e^2 [2 \pi]^4 \frac{1}{ 2 q_{i,0} } \frac{1}{ [2 \pi L]^2}   \frac{d^3\pmb{q}_f}{ 2 q_{f,0} } \frac{d^3\pmb{k}_f}{ 2 \omega_f } \sum_{n = - \infty}^{\infty} \int_{- \pi L}^{ \pi L} d\varphi \int_{- \pi L}^{ \pi L} d\vartheta\
\delta^{(4)}\lc q_i + s_n  k - q_f - k_f \rc  \\
&\times \bigg\{    16 m^2 - 8q_i \cdot q_f - 4 e^2 \mathfrak{a}^2 \lL \frac{k \cdot q_i}{k \cdot q_f}  + \frac{k \cdot q_f}{k \cdot q_i} \rL
- 4 e^2 \lL \gA^2_\varphi + \gA^2_\vartheta - \gA_\varphi \cdot \gA_\vartheta \lL  \frac{k \cdot q_i}{k \cdot q_f} + \frac{k \cdot q_f}{k \cdot q_i}  \rL \rL \\
&- 4 e \lL \gA_\varphi + \gA_\vartheta \rL \cdot \lL q_i \lL \frac{k \cdot q_f}{k \cdot q_i} - 1  \rL + q_f \lL \frac{k \cdot q_i}{k \cdot q_f} - 1  \rL  \rL 
\bigg\}
\exp \lc i \lL \frac{q_f \cdot \mathcal{B}_{(\varphi,\vartheta)} }{k \cdot q_f}  - \frac{q_i \cdot \mathcal{B}_{(\varphi,\vartheta)} }{k \cdot q_i} \rL  \rc\\
&\times \exp \lc i s_n [\varphi - \vartheta] \rc.
\label{eq_fin-dW}
} 
Note that the following short notation
\eqn{
\mathcal{B}^\mu_{(\varphi,\vartheta)} \equiv e \int_{\vartheta}^{\varphi} dz\ \gA^\mu(z)
\label{eq_defBcal}
}
has been introduced.
%

\subsection{Phase-space in lepton-photon collision}
\label{subsec_phocs}

%
Choosing a coordinate system where
the external field photon momentum is oriented in $\hat z$-direction, i.e.\
$k^\mu = [\omega,0,0,\omega]$, allows to simplify the subsequent phase-space integration when calculating the total probability $W_{cs}$ for the scattering process.
This special coordinate system we call photon-oriented coordinate system (POCS).
From Eq.~\eqref{eq_fin-dW} one can read that the Lorentz invariant phase-space measure ($d\text{LIPS}$) is given by
\eqn{
\int d \text{LIPS}\ \delta^{(4)}\lc q_i + s_n k - q_f - k_f \rc = \int \frac{d^3 \pmb{q}_f}{2 q_{f,0}}  \int \frac{d^3 \pmb{k}_f}{2 \omega_f}  \delta^{(4)}\lc q_i + s_n  k - q_f - k_f \rc.
}
Defining a Lorentz invariant integration variable, similar to Ref.~\cite{landau},
\eqn{
u _{cs}\equiv \frac{k \cdot k_f}{k \cdot q_f}
\label{eq_invariant-u}
}
simplifies the phase-space integration in a crucial way for specific field configurations, 
as we will experience in forthcoming calculations.
The new variable $u_{cs}$ presents just the angular ratio between the two planes $k \cdot k_f$ and $k \cdot q_f$.
In addition to this invariant variable, we 
apply for each term in the series (denoted by subscript $n$) the corresponding center-of-mass system (CMS). 
This mode-independent $\text{CMS}_{n,cs}$ frame is characterised by setting\footnote{\label{fnote_contradiction}In the literature often the assumption $\pmb{k} \parallel \hat z$ is taken in addition to the constraint that the initial electron is supposed to be at rest at average, i.e. $\pmb{q}_i = \pmb{0}$. 
This condition leads for instance in the POCS-$\text{CMS}_{n,cs}$ frame, going to be constructed in the present section, to $s_n \omega = 0$, which obviously cannot be fulfilled for any arbitrary field frequency $\omega$, because $s_n \geq 1$ and $\omega > 0$ have to be fulfilled due to momentum conservation, App.~\ref{app_cs-spl}. 
Only in case of imposing in addition the assumption of soft-energetic photons, the condition $\pmb{q}_i = \pmb{0}$ does not lead to a contradiction, see Ref.~\cite{landau}.
More details regarding this approximation can be found in App.~\ref{app_cs-spl}. 
However, in our case,  
we do not impose further constraints on the initial effective electron momentum $q_i$, but keep it of general form as being on-shell with respect to the previously introduced quasi-mass, i.e. $q_i^2 = m_{\star}^2$.}
\eqn{
s_n \pmb{k} + \pmb{q}_i = \pmb{k}_f + \pmb{q}_f = 0.
\label{eq_cms}
}
As an immediate consequence, Eq.~\eqref{eq_cms} will force the initial electron momentum to obey $q_{i,1} = q_{i,2} = 0$, such that $\pmb{q}_i = [0,0,q_{i,3}]$ applies in most general case for the $\text{CMS}_{n,cs}$ frame.
With this choice, one gets
\eqnsplit{ 
&\gA \cdot q_i = - \gA_{,3} q_{i,3},\\
&\mathcal{B}_{(\varphi,\vartheta)} \cdot q_i = - \mathcal{B}_{(\varphi,\vartheta),3} q_{i,3}.
\label{eq_gauge-extension-cs}
}
Note that the $\text{CMS}_{n,cs}$ provides a CMS frame on ``perturbative'' level for each of the  harmonic modes for the bosonic field quanta, characterised by the summation index $n$. In principle this would mean that for different parameter settings the $\text{CMS}_{n,cs}$ frame has to be changed according to the dominating contribution in the series for a specific $n$.

Eq.~\eqref{eq_cms}  leads immediately to $|\pmb{q}_f|  = |\pmb{k}_f|$. Notice that the final photon energy is naturally expected to be $\omega_f > 0$ and final electron and photon have obviously opposite momentum directions due to $\text{CMS}_{n,cs}$. 
The total energy of the system reads as
\eqn{
\epsilon_{n,cs} \equiv s_n \omega + q_{i,0} = \omega_f + q_{f,0}. 
}
%
With taking this into account, we rewrite the delta function in following way
\eqn{
\delta^{(4)}\lc q_i + s_n k - q_f - k_f \rc \longrightarrow \delta \lc \epsilon_{n,cs} - \omega_f - q_{f,0}\rc \delta^{(3)} \lc \pmb{k}_f + \pmb{q}_f \rc
}
and parametrize the phase-space by using spherical coordinates for the final effective electron momentum, i.e. 
$|\pmb{q}_f|$, $\phi$ and $\theta$.  
After performing some algebraic operations we conclude 
\eqn{
\int d\text{LIPS}\ \delta^{(4)}\lc q_i + s_n k - q_f - k_f \rc \longrightarrow   
\frac{1}{ [2 \pi]^4}  \int_{0}^{2 \pi}   d\phi  \int_{ -1}^{1}  \frac{|\pmb{q}_f|}{ 8 \epsilon_{n,cs}} d\cos(\theta).
\label{eq_dLIPS-trans}
}
Here $\theta$ is the polar angle between $\pmb{q}_f$ and the $\hat z$-axis. 
Due to POCS where $\pmb{k} \parallel \hat z$, $\theta$ becomes identical to the angle between $\pmb{q}_f$ and $\pmb{k}$, i.e. $\cos(\theta) = \frac{\pmb{k} \cdot \pmb{q}_f}{|\pmb{k}| |\pmb{q}_f|}$ and one can
express the differential $d\cos(\theta)$ in terms of $d u_{cs}$. Taking the definition of $u_{cs}$ from Eq.~\eqref{eq_invariant-u}, we obtain
\eqn{
u_{cs} + 1 = \frac{\epsilon_{n,cs} } { q_{f,0} - |\pmb{q}_f| \cos(\theta) }
} 
and thus accordingly
\eqn{
d\cos(\theta) = \frac{\epsilon_{n,cs} }{|\pmb{q}_f|[1+u_{cs} ]^2} d u_{cs}.
\label{eq_dcosTheta}
}
Applying Eq.~\eqref{eq_dcosTheta} to \eqref{eq_dLIPS-trans}, we end up for Compton scattering in our chosen POCS-$\text{CMS}_{n,cs}$ system 
with the following reduced Lorentz-invariant phase-space measure, denoted by $d\widehat{\text{LIPS}}_{cs}$,
\eqnsplit{
\int_{0}^{2 \pi}   d\phi  \int_{ -1}^{1}  \frac{|\pmb{q}_f|}{ 8 \epsilon_{n,cs}} d\cos(\theta) = \int d\widehat{\text{LIPS}}_{cs}
\equiv \int_{0}^{2 \pi} d\phi \int_{0}^{ u_{n,cs} } \frac{d u_{cs}}{2^3 [1+u_{cs}]^2 },
\label{eq_reduced-ps}
}
and the upper and lower integration limit, respectively, read as
\begin{align}
\max\{u_{cs}\} &\longrightarrow  u_{n,cs} \equiv \frac{\epsilon_{n,cs}^2}{m_{\star}^2} - 1 \quad  [ \theta = 0 ], \label{eq_int-up-limit}\\
\min\{u_{cs}\} &\longrightarrow 0 \quad [ \theta = \pi ]. \label{eq_int-low-limit}
\end{align}
In the following we will be using this POCS-$\text{CMS}_{n,cs}$ frame, characterised by the reduced phase-space measure $\int d\widehat{\text{LIPS}}_{cs}$, for calculating 
the general total scattering probability for a generalised class of plane-wave background fields.

\subsection{Total scattering probability}
\label{subsec_Wcs}

We calculate the total scattering probability via applying Eq.~\eqref{eq_reduced-ps} on \eqref{eq_fin-dW}.
Note that due to energy and momentum conservation in POCS-$\text{CMS}_{n,cs}$, the original index $n$ for the total sum in expression \eqref{eq_fin-dW} has to obey $n \geq 1$. Therefore, we get rid of all summations over negative $n$ in the series and conclude accordingly:
\eqnsplit{
&W_{cs} =  \frac{  e^2  }{ [2 q_{i,0}]   } \frac{1}{[2 \pi L]^2}    \sum_{n = 1}^{\infty} \int_{- \pi L}^{ \pi L} d\varphi \int_{- \pi L}^{ \pi L} d\vartheta  \exp \lc i s_n [\varphi - \vartheta] \rc \int d\widehat{\text{LIPS}}_{cs}\\
&\times \bigg\{ 
16 m^2 - 8 q_i \cdot q_f - 4 e^2 \mathfrak{a}^2 \lL \frac{1+[1+u_{cs}]^2}{1+u_{cs}} \rL
- 4 e \lL \gA_\varphi + \gA_\vartheta \rL \cdot \lL q_i \lL \frac{1}{1 + u_{cs}} - 1  \rL + q_f u_{cs}  \rL\\
&- 4 e^2 \lL \gA^2_\varphi + \gA^2_\vartheta - \gA_\varphi \cdot \gA_\vartheta \lL  \frac{1+[1+u_{cs}]^2}{1+u_{cs}} \rL  \rL 
\bigg\}
\exp \lc i \lL [1+u_{cs}] \frac{ q_f \cdot \mathcal{B}_{(\varphi,\vartheta)} }{k \cdot q_i}  - \frac{q_i \cdot \mathcal{B}_{(\varphi,\vartheta)} }{k \cdot q_i} \rL  \rc,
\label{eq_totW1}
}
where the invariant variable $u_{cs}$ has been employed appropriately to establish the relation
$ \frac{k \cdot q_i}{k \cdot q_f} = 1 + u_{cs}$.

In the remaining part of this section we will derive a final general integral expression for the scattering probability where only the corresponding field shape related integrations over the Fourier phases$\int d\varphi$, $\int d\vartheta$ and the final reduced phase-space integral $\int du_{cs}$ shall be kept. Consequently, we have to perform the integration over the azimuth phase-space angle $\phi$ as a next step. Regarding this we will proceed in the following way.

Without loss of generality, we choose $a_{1,0} = a_{2,0} = 0$ for the field polarisation vectors in \eqref{eq_gen-field}, which leads immediately to
$\mathcal{A}_{\varphi,0} = \mathcal{A}_{\vartheta,0} = 0$
and
$\mathcal{B}_{(\varphi,\vartheta),0} =  0$, respectively.
As a next step, we define for avoiding redundant complications new $\phi$-independent expressions $\mathfrak{T}_j$ and $\mathfrak{B}_j$. Those are presented in App.~\ref{app_definitions}, Eqs.~\eqref{def_B1} - \eqref{def_T3}. Due to their lengthy shape, we do not present them here in great detail.
By using these definitions we obtain the simplified expression
\eqnsplit{
W_{cs} &= e^2  \frac{1}{ [2 q_{i,0}] } \frac{1}{ [2 \pi L]^2 } \frac{1}{2^3} 
\sum_{n = 1}^{\infty} \int_{-\pi L}^{\pi L} d\varphi \int_{-\pi L}^{\pi L} d\vartheta \exp\lc i s_n [\varphi - \vartheta] \rc\\
&\times \int_{u = 0}^{u_{n,cs} }  \frac{d u_{cs} }{[1+u_{cs} ]^2}  \int_{\phi = 0}^{2 \pi}   d\phi
\lL \mathfrak{T}_3 + \mathfrak{T}_2\sin(\phi) + \mathfrak{T}_1\cos(\phi)  \rL
\exp\lc -i \mathfrak{B}_3 \rc\\
&\times \exp\lc - i \lL \mathfrak{B}_1 \cos(\phi) + \mathfrak{B}_2 \sin(\phi) \rL \rc.
\label{eq_totW2}
}
In order to integrate over the azimuth angle $\phi$ , one has to solve the following definite integrals 
\begin{align}
&\int_{0}^{2\pi}  \exp\lc - i \mathfrak{B}_1 \cos(\phi) - i \mathfrak{B}_2 \sin(\phi) \rc\ d\phi,    \label{eq_1st-int}\\
&\int_{0}^{2\pi} \mathfrak{T}_2 \sin(\phi) \exp\lc - i \mathfrak{B}_1 \cos(\phi) - i \mathfrak{B}_2 \sin(\phi) \rc\ d\phi,    \label{eq_2nd-int}\\
&\int_{0}^{2\pi} \mathfrak{T}_1 \cos(\phi) \exp\lc - i \mathfrak{B}_1 \cos(\phi) - i \mathfrak{B}_2 \sin(\phi) \rc\ d\phi.  \label{eq_3rd-int}
\end{align} 
For integral \eqref{eq_1st-int} we obtain directly a solution of the form:
\eqnsplit{
\int_{0}^{2\pi}  \exp\lc - i \mathfrak{B}_1 \cos(\phi) - i \mathfrak{B}_2 \sin(\phi) \rc\ d\phi =
2 \pi \mathcal{J}_0 \lc \sqrt{ \mathfrak{B}_1^2 + \mathfrak{B}_2^2 } \rc,
\label{eq_res-1st-int}
}
where $\mathcal{J}_0$ denotes the standard Bessel function of the first kind and zeroth order. 

Concerning the integrals \eqref{eq_2nd-int} and \eqref{eq_3rd-int}, we write the corresponding pre-factors $\mathfrak{T}_2 \sin(\phi)$ and $\mathfrak{T}_1 \cos(\phi)$, respectively, as exponential functions and solve the resulting integrals. Linearisation afterwards with respect to $\mathfrak{T}_1$ and $\mathfrak{T}_2$, respectively, leads to the correct results. 
We obtained analytically closed solutions in terms of standard Bessel functions\footnote{Note that particularly $\mathcal{J}_{2 s}$ are axial-symmetric functions $\forall\ s \in \mathbb{N}_0$, i.e. $\mathcal{J}_{2 s}(Z) = \mathcal{J}_{2 s}(-Z)$.}  $\mathcal{J}_0$ and $\mathcal{J}_2$
and the final solution for the integration over $\phi$
in Eq.~\eqref{eq_totW2}
becomes consequently
\eqnsplit{
2 \pi  \bigg\{ \mathfrak{T}_3  \mathcal{J}_0 \lc  \sqrt{  \mathfrak{B}_1^2 + \mathfrak{B}_2^2  } \rc
- \frac{ i  \lL \mathfrak{B}_2 \mathfrak{T}_2 +  \mathfrak{B}_1 \mathfrak{T}_1 \rL  }{ 2 } 
\lL  \mathcal{J}_0 \lc  \sqrt{  \mathfrak{B}_1^2 + \mathfrak{B}_2^2  } \rc 
+ \mathcal{J}_2 \lc  \sqrt{  \mathfrak{B}_1^2 + \mathfrak{B}_2^2  } \rc \rL \bigg\}.
\label{eq_phi-int-res}
}
The total scattering probability in the case of a generally shaped plane-wave background field, i.e. some arbitrary form of the functions ${A_1}_r$ and ${A_2}_r$ from Eq.~\eqref{eq_gen-field} $\forall\ r \in \{ 1,\ldots, N \}$, is therefore given by:
\begin{framed}
\eqnsplit{
&W_{cs} = \frac{e^2  [2\pi]  }{2^3 [2 q_{i,0}] [2 \pi L]^2} \sum_{n = 1}^{ \infty} \int_{-\pi L}^{\pi L} d\varphi \int_{-\pi L}^{ \pi L}d\vartheta \exp\lc i s_n [\varphi - \vartheta] \rc 
\int_{0}^{u_{n,cs}}  \frac{d u_{cs} }{[1+u_{cs}]^2} \\
&\times  
\bigg\{
\mathfrak{T}_3 \mathcal{J}_0 \lc  \sqrt{  \mathfrak{B}_1^2 + \mathfrak{B}_2^2 } \rc  
- \frac{ i \lL \mathfrak{B}_2 \mathfrak{T}_2  +  \mathfrak{B}_1 \mathfrak{T}_1  \rL }{ 2 } \lL \mathcal{J}_0 \lc  \sqrt{  \mathfrak{B}_1^2 + \mathfrak{B}_2^2  } \rc 
+ \mathcal{J}_2 \lc  \sqrt{  \mathfrak{B}_1^2 + \mathfrak{B}_2^2  } \rc   \rL
\bigg\}\\
&\times \exp \lc -i \mathfrak{B}_3 \rc.
\label{eq_gen-fin-totW-scattering}
}
\end{framed}
Expression \eqref{eq_gen-fin-totW-scattering} can be regarded as the most general form for the total scattering probability in the POCS-$\text{CMS}_{n,cs}$ frame.
One should note that all definitions which occur in Eq.~\eqref{eq_gen-fin-totW-scattering} and listed in App.~\ref{app_definitions} can be rewritten in POCS-$\text{CMS}_{n,cs}$ without the remaining $\theta$-dependencies
according to
$\cos(\theta) = \frac{1}{1+u_{cs}} \lL \frac{u_{cs} q_{f,0} }{\omega_f}  - 1 \rL$.
As a last step, one can re-express the final electron and photon energies applying Eqs.~\eqref{eq_effm_2} and \eqref{eq_int-up-limit} 
in terms of dimensionless parameters $\xi$ and $ u_{n,cs} $ as follows: 
\begin{align}
\omega_f &= \frac{\epsilon_{n,cs}^2 - m_{\star}^2}{2 \epsilon_{n,cs} } = \frac{ m_{\star} }{2} \frac{u_{n,cs}}{\sqrt{1 + u_{n,cs}}}, \label{eq_wf-circ-zprop}\\
q_{f,0} &=  \frac{\epsilon_{n,cs}^2 + m_{\star}^2}{2 \epsilon_{n,cs} } = \frac{ m_{\star} }{2} \frac{2 + u_{n,cs}}{\sqrt{1 + u_{n,cs}}}. \label{eq_qf0-circ-zprop}
\end{align}
Let us remind that the first equalities in both lines are directly resulting from the $\text{CMS}_{n,cs}$. Combining Eq.~\eqref{eq_wf-circ-zprop} with~\eqref{eq_qf0-circ-zprop} leads to the useful relation
\eqn{
\frac{q_{f,0}}{\omega_f} = \frac{u_{n,cs} + 2}{u_{n,cs}},
\label{eq_energy-ratio}
}
which will be discussed in App.~\ref{app_cs-spl}
with respect to the comparison with the SPL approach.
Eqs.~\eqref{eq_wf-circ-zprop} - \eqref{eq_energy-ratio} yield consequently the following replacements for the angular parts in terms of dimensionless parameters $ u_{cs}, u_{n,cs}$, cf. Eqs.~\eqref{eq_invariant-u} and \eqref{eq_int-up-limit}:
\begin{align}
\cos(\theta) &= \frac{ 2 u_{cs} - u_{n,cs} + u_{cs} u_{n,cs} }{ u_{n,cs} [1 + u_{cs}]  }, \label{eq_cosTheta-dep}\\
\sin(\theta) &= \frac{2}{[1 + u_{cs}] u_{n,cs}} \sqrt{ u_{cs}  [1 + u_{n,cs}] [u_{n,cs}  - u_{cs}]  } \label{eq_sinTheta-dep}.
\end{align}
%

%

\section{Compton scattering in head-on lepton-photon collisions}
\label{sec_compton_nc}

\subsection{Closed analytical result}

This section shall deal with the generalisation of Compton scattering, schemed in Fig.~\ref{fig_compton}, in the environment 
of a laser-like background field. 
In the following we therefore apply a specific field configuration and evaluate and simplify each term in \eqref{eq_gen-fin-totW-scattering}, respectively.
This field choice has been applied before, cf. i.e. Refs.~\cite{landau,ritus}, assuming the in-coming particle at rest and considering the photons in SPL, cf. App.~\ref{app_cs-spl}. Note, however, that latter constraint on the photon energies is a direct consequence of the initial particle rest-frame in the POCS-$\text{CMS}_{n,cs}$ approach. Imposing only small energies, $s_n \omega \approx 0 $, is strongly required 
to resolve 
the previously mentioned contradiction, see footnote~\ref{fnote_contradiction}.
Within the present discussion, however, we will avoid some further energetic constraints on the photon energies. For completeness, a detailed calculation in SPL is also presented in App.~\ref{app_cs-spl}.
The transition to SPL may be seen therefore as a confirmation of the general total scattering probability for a class of generalised plane-wave fields presented in Eq.~\eqref{eq_gen-fin-totW-scattering}, and the quite well-known perturbative solution, often used in the literature, has been reproduced starting from this general formula.

In the following we consider a more general case:
the electron has a momentum in $\hat z$-direction. 
The external field is circularly polarised in the $(\hat x,\hat y$)-plane, whereas the photon wave vector is oriented in $\hat z$-direction, cf. the previously constructed POCS-$\text{CMS}_{n,cs}$ in Sec.~\ref{subsec_phocs}. Obviously, this will provide a suitable choice for electromagnetic fields of accessible lab laser beams. The field polarisation vectors are chosen $\pmb{a}_1 \parallel \hat x$, $\pmb{a}_2 \parallel \hat y$
and the final external field shape is given by
\eqn{
\gA^\mu(\eta) = a_1^\mu \cos \eta  + a_2^\mu \sin \eta,
\label{eq_circ-field}
}
which has the consequence that according to Eq.~\eqref{eq_A2-constraint} the equality $a_i \cdot a_j \equiv \mathfrak{a}^2 \delta_{ij}$ has to be imposed. 
Due to the $2 \pi$-periodicity of $\gA^\mu(\eta)$
we take $L = 1$, so that $s_n = n$.

\subsubsection{Separating the Fourier phases in pre-factors}

By considering these assumptions, we deduce the following components of Eq.~\eqref{eq_defBcal}
\eqnsplit{
\mathcal{B}_{(\varphi,\vartheta),0} &=\mathcal{B}_{(\varphi,\vartheta),3} = 0,\\
\mathcal{B}_{(\varphi,\vartheta),1} &= e a_{1,1} \lL \sin\varphi - \sin\vartheta  \rL,\\
\mathcal{B}_{(\varphi,\vartheta),2} &= e a_{2,2}\lL \cos\vartheta - \cos\varphi  \rL.
}
Note that, since this is the complete energetic non-constrained case where electron and external field photons collide along the $z$-axis, we label the relevant expressions below with a subscript $\mathrm{coll}$
and separate
the field phase dependent parts for all definitions. Applying the field choice from above, we get, cf. Eqs.~\eqref{def_B1} - \eqref{def_T3},
\begin{align}
&\mathfrak{B}_3  \longrightarrow 0, \label{weq_gen-circ-zprop_b34}\\
&\mathfrak{B}_{1} \mathfrak{T}_{1} + \mathfrak{B}_{2} \mathfrak{T}_{2} \longrightarrow \mathfrak{R}_{21,\mathrm{coll}} \sin(\varphi - \vartheta), \label{eq_gen-circ-zprop_b2t49plusb2t38}\\
&\mathfrak{T}_{3} \longrightarrow \mathfrak{T}_{31,\mathrm{coll}} + \mathfrak{T}_{32,\mathrm{coll}} \sin^2 \lc \frac{\varphi - \vartheta}{2} \rc, \label{eq_gen-circ-zprop_t00}
\end{align}
with
\begin{align}
\mathfrak{R}_{21,\mathrm{coll}} 
&\equiv 
\frac{  8 m^4 \xi^2 u_{cs}^2 [1 + \xi^2] [ u_{n,cs} - u_{cs} ] }{ k \cdot q_i  [1 + u_{cs}] },
\label{eq_gen-circ-zprop_b21}\\
\mathfrak{T}_{31,\mathrm{coll}} 
&\equiv 
16 m^2 + 8 m^2 \xi^2
- \frac{  8 m \sqrt{1 + \xi^2} [2 + u_{n,cs} ]  q_{i,0}  }{2 \sqrt{1 + u_{n,cs}} }\\
&+ \frac{ 8 m \sqrt{1 + \xi^2} [2 u_{cs} - u_{n,cs} + u_{cs} u_{n,cs}]  q_{i,3}  }{ 2 [1 + u_{cs}] \sqrt{1 + u_{n,cs}} } , \label{eq_gen-circ-zprop_t01}\\
\mathfrak{T}_{32,\mathrm{coll}} 
&\equiv
8 m^2 \xi^2   \lL \frac{1 + [1+u_{cs}]^2}{1+u_{cs}} \rL.
\label{eq_gen-circ-zprop_t02}
\end{align}
Note that final energies $\omega_f$ and $q_{f,0}$, Eqs.~\eqref{eq_wf-circ-zprop} and \eqref{eq_qf0-circ-zprop}, are expressed completely in terms of the integration variable $u_{cs}$, its upper limit $u_{n,cs}$\footnote{Note that $u_{n,cs}$ remains unchanged since no constraint, cf. footnote \ref{fnote_contradiction}, on the final energies is imposed.} and the intensity parameter $\xi$, cf. also \eqref{eq_effm_2}. Polar angle-dependent terms, cf. \eqref{def_B1} - \eqref{def_T3}, are replaced using Eqs.~\eqref{eq_cosTheta-dep} and \eqref{eq_sinTheta-dep}. 

\subsubsection{Separating the Fourier phases in function arguments}

The argument $\sqrt{ \mathfrak{B}_1^1 + \mathfrak{B}_2^2 }$ of all appearing Bessel functions in the general expression $W_{cs}$, Eq.~\eqref{eq_gen-fin-totW-scattering}, can be written as:
\eqn{
\sqrt{ \mathfrak{B}^2_{1} + \mathfrak{B}^2_{2} } \longrightarrow  2 Z_{n,cs}  \sqrt{  \sin^2 \lc \frac{\varphi - \vartheta}{2} \rc },
\label{eq_B1B2cs}
}
where we have introduced
\eqn{
Z_{n,cs} \equiv \frac{ m^2  \xi  \sqrt{1+\xi^2} }{ \omega m_{\star} \sqrt{1 + u_{n,cs}} } \sqrt{ u_{cs} u_{n,cs} - u_{cs}^2 }
\label{eq_Z-circ-zprop}
}
using Eqs.~\eqref{eq_wf-circ-zprop}, \eqref{eq_cosTheta-dep} and $k \cdot q_i = \omega m_{\star} \sqrt{1 + u_{n,cs}}$ in POCS-$\text{CMS}_{n,cs}$ according to \eqref{eq_int-up-limit}. 
As derived in App.~\ref{app_cs-spl}, the SPL case will lead to a small upper integration limit and $Z_{n,\mathrm{soft}}$ including modified functions $\mathfrak{R}_{21,\mathrm{soft}}$ and $\mathfrak{T}_{31,\mathrm{soft}}$, respectively.

For the following discussion let us set $q_{i,0} \equiv \lambda m_{\star}$ with the often used parameter $\lambda > 1$, leading to $q_{i,3} = - m_{\star} \sqrt{ \lambda^2 - 1 } < 0$. Taking $n \omega = - q_{i,3}$ in POCS-$\text{CMS}_{n,cs}$ one gets 
\eqn{
\lambda = \sqrt{ \frac{u_{n,\mathrm{soft}}^2}{ 4 } + 1 }
\label{eq_lambda}
} 
and
\eqn{
q_{i,3} = \frac{- m_{\star} u_{n,\mathrm{soft}} }{2},
\label{eq_qi3-in-unsoft}
}
where $u_{n,\mathrm{soft}} = 2 n \omega / m_{\star} $, see App.~\ref{app_cs-spl}.
This gives for the corresponding upper integration limit in the present, still unconstrained, case:
\eqn{
u_{n,cs} \longrightarrow u_{n,\mathrm{coll}} \equiv \lL \frac{u_{n,\mathrm{soft}}}{2} + \lambda \rL^2 - 1.
}
Consequently, we derive for \eqref{eq_Z-circ-zprop}, using Eq.~\eqref{eq_qi3-in-unsoft}
\eqn{
Z_{n,cs} \longrightarrow Z_{n,\mathrm{coll}}  =
2 n \frac{ \xi }{ \sqrt{1 + \xi^2} } 
\sqrt{ \frac{  u_{cs} u_{n,\mathrm{coll}} - u_{cs}^2  }{ [1 + u_{n,\mathrm{coll}}] u_{n,\mathrm{soft}}^2   } }.
\label{eq_Z-ncz-final}
}
The limiting value $\lambda \rightarrow 1$ will directly lead to the initial rest-frame, i.e. $q_{i,3} \rightarrow 0$, according to Eqs. \eqref{eq_lambda} and \eqref{eq_qi3-in-unsoft}.
Considering this limit will therefore directly lead to the SPL approach, i.e. $n \omega \approx 0$. Contrary, if one expects that the initial electron has non-vanishing momentum, we have to choose the parameters, cf. Eq.~\eqref{eq_lambda}, such that 
\eqn{
\frac{n^2 \omega^2 }{ m_{\star}^2 } \gg 0
\label{eq_ncz-condition}
} 
or
\eqn{
u_{n,\mathrm{soft}} \gg 0,
\label{eq_ncz-condition2}
}
respectively.
Note that otherwise $\mathfrak{R}_{21,\mathrm{coll}}$ will vanish completely, see App.~\ref{app_cs-spl}.
Because of that, even for $\xi \ll 1$ and $\omega \ll m$ the present non-constrained result will be relevant for sufficient large $n$. Or say, for a proper ``correction'' in the multi-photon-regime one has to consider the generalised solution \eqref{eq_gen-fin-totW-scattering} even for small field parameters since the total amount of energy for the absorbed field quanta will be sufficiently large although each of them is supposed to be soft.

Note that due to the equality $n \omega + q_{i,3} = 0$
the particle momentum is being refined for every single mode in the perturbation series since $n$ varies. This has simply the consequence that by fixing the initial electron momentum, it will be possible to single out the contribution coming from one specific harmonic mode in the series. This provides actually an interesting possibility of adjusting the measurement sensitivity according to any perturbation level: i.e. the dominant term for a specific parameter combination in the series will provide the electron momentum for the ``ideal''  $\text{CMS}_{n,cs}$ frame, respectively. Moreover, a proper lab CMS is therefore expected to be approximately identical to $\text{CMS}_{n,cs}$ of the dominant contribution.

\subsubsection{Integrations over all Fourier phases}

Applying Eqs.~\eqref{eq_B1B2cs}, \eqref{eq_Z-circ-zprop} and \eqref{weq_gen-circ-zprop_b34} - \eqref{eq_gen-circ-zprop_t00}  to Eq.~\eqref{eq_gen-fin-totW-scattering}  yields following total scattering probability:
\eqnsplit{
&W_{cs,\mathrm{coll}} \equiv \frac{e^2    }{2^3 [2 q_{i,0}]  [2 \pi]  } 
\sum_{n = 1}^{ \infty} 
\int_{-\pi}^{ \pi} d\varphi 
\int_{-\pi}^{ \pi} d\vartheta\ 
\exp\lc i n [\varphi - \vartheta] \rc 
\int_{0}^{u_{n,\mathrm{coll}}}  \frac{du_{cs}}{[1+u_{cs}]^2} \\
&\times \bigg\{
\mathcal{J}_0 \lc 2 Z_{n,\mathrm{coll}} \sin \lc \frac{\varphi - \vartheta}{2} \rc \rc \lL \mathfrak{T}_{31,\mathrm{coll}} + \mathfrak{T}_{32,\mathrm{coll}} \sin^2 \lc \frac{\varphi - \vartheta}{2} \rc  \rL
- \lL \frac{ i \mathfrak{R}_{21,\mathrm{coll}} \sin(\varphi - \vartheta) }{ 2 } \rL\\
&\times \lL \mathcal{J}_0 \lc  2 Z_{n,\mathrm{coll}} \sin \lc \frac{\varphi - \vartheta}{2} \rc   \rc
+ \mathcal{J}_2 \lc  2 Z_{n,\mathrm{coll}}  \sin \lc \frac{\varphi - \vartheta}{2} \rc   \rc \rL
\bigg\}.
\label{eq_totW-circ-zprop-1}
}  
Since the Fourier phases $\varphi$ and $\vartheta$ only occur in the constellation $\varphi - \vartheta$, it is useful to change to
\eqn{
d\varphi d\vartheta = - \frac{1}{2} d\varphi^- d\varphi^+,
\label{eq_variable-change}
}
where obviously the light-cone coordinates have been chosen:
\begin{align}
\varphi^-  &\equiv \frac{\varphi - \vartheta}{2}, \label{eq_phi-minus}\\
\varphi^+ &\equiv \frac{\varphi + \vartheta}{2}, \label{eq_phi-plus}
\end{align}
where the integration about $\varphi^+$ gives
$\int_{-\pi}^{\pi} d\varphi^+ = 2 \pi$.
Using the Euler formula, we can apply Neumann's integral identity\footnote{Sec. 2.6 in \cite{watson}} relating
Bessel functions $\mathcal{J}_0$ and $\mathcal{J}_n\ \forall\ n \in \mathbb{N}$:
\eqn{
2 \pi \mathcal{J}_n^2 (Z) =  \int_{-\pi}^{\pi} dz\ \mathcal{J}_0 \lc2 Z \sin(z) \rc \exp\lc 2 i n z \rc
\label{eq_neumann-id}
}
$\forall Z \in \mathbb{R}$. Applying Eq.~\eqref{eq_neumann-id} to Eq.~\eqref{eq_totW-circ-zprop-1} we obtain
\eqnsplit{
W_{cs,\mathrm{coll}} &= \frac{e^2    }{2^3 [2 q_{i,0}]  2  } 
\sum_{n = 1}^{\infty} 
\int_{0}^{u_{n,\mathrm{coll}}}  \frac{du_{cs}}{[1+u_{cs}]^2}
\bigg\{
- 2 \pi \mathfrak{T}_{31,\mathrm{coll}} \mathcal{J}_n^2(Z_{n,\mathrm{coll}})\\
&+ \frac{  2 \pi  \mathfrak{T}_{32,\mathrm{coll} }}{4} \lL  \mathcal{J}_{n-1}^2(Z_{n,\mathrm{coll}}) + \mathcal{J}_{n+1}^2(Z_{n,\mathrm{coll}})  - 2 \mathcal{J}_n^2(Z_{n,\mathrm{coll}}) \rL\\
&+ \frac{  2 \pi   \mathfrak{R}_{21,\mathrm{coll} } }{4}   \lL \mathcal{J}_{n-1}^2(Z_{n,\mathrm{coll}})  - \mathcal{J}_{n+1}^2(Z_{n,\mathrm{coll}})  \rL\\
&+ \frac{\mathfrak{R}_{21,\mathrm{coll} }}{4} \int_{-\pi}^{\pi} d\varphi^-   \mathcal{J}_2 \lc 2 Z_{n,\mathrm{coll}} \sin\varphi^-  \rc
\lL e^{2 i [n-1] \varphi^-} - e^{2 i [n+1] \varphi^-}  \rL
\bigg\}.
\label{eq_totW-circ-zprop-4}
}
It is evident that such non-restricted energy contributions lead to additional terms (proportional to $\mathfrak{R}_{21,\mathrm{coll}}$), extending substantially the SPL approximation.
One still has the integral over $\varphi^-$ in Eq.~\eqref{eq_totW-circ-zprop-4}, which has the general form
$\int_{-\pi}^{\pi} dx\ \mathcal{J}_2 \lc 2 Z \sin(x)  \rc e^{2 i n x}$
with $n \in \mathbb{Z}$. This expression differs in a crucial way from the one in~\eqref{eq_neumann-id}. Due to the replacement of $\mathcal{J}_0$ by $\mathcal{J}_2$, the previously mentioned Neumann's integral identity cannot be applied directly. 
We therefore use Sonine's integral \cite{watson} and replace $\mathcal{J}_2$ in terms of $\mathcal{J}_0$.
Applying in addition Neumann's integral identity \eqref{eq_neumann-id} yields:
\eqnsplit{
&\int_{-\pi}^{ \pi} d\varphi\ \mathcal{J}_2 \lc 2 Z \sin\varphi  \rc \exp(2 i n \varphi) \longrightarrow
\pi Z^2  \int_{0}^{\pi/2} dx\  \sin(x) \cos^3(x)\\
&\times \lL 2 \mathcal{J}_n^2 (Z \sin(x))  - \mathcal{J}_{n-1}^2 (Z \sin(x))  - \mathcal{J}_{n+1}^2 (Z \sin(x)) \rL.
\label{eq_sonine-neumann-res}
}
By performing a variable transformation we achieve moreover
\eqnsplit{
&\int_{-\pi}^{ \pi} d\varphi\ \mathcal{J}_2 \lc 2 Z \sin\varphi  \rc \exp(2 i n \varphi)\\ 
&= \pi Z^2 \int_{0}^{1} dy\ y(1 - y^2) \lL  2 \mathcal{J}_{n}^2 (Z y)
- \mathcal{J}_{n-1}^2 (Z y)  - \mathcal{J}_{n+1}^2 (Z y) \rL\\
&= 2 \pi \lL  \Theta\lc \frac{1}{2} -  |n| \rc \mathcal{J}_{n+1}^2(Z) - \Theta\lc n - \frac{1}{2} \rc \mathfrak{F}_n(Z) \rL,
\label{eq_res-mod-Neu-ohneNegOne}
} 
where $\Theta$ denotes the usual Heaviside step function\footnote{$\Theta(x) = \begin{cases} 0\quad \forall\ x < 0\\ 1\quad \forall\ x> 0 \end{cases}$}. 
The function $\mathfrak{F}_n(Z)$ is defined in App.~\ref{app_definitions} in terms of generalised Hypergeometric functions and the usual Gamma function.
Note that latter result \eqref{eq_res-mod-Neu-ohneNegOne} is only valid $\forall\ n \geq 0$ and $Z \geq 0$.
For the remaining integral in Eq.~\eqref{eq_totW-circ-zprop-4} we therefore conclude following solution
\eqnsplit{
&\int_{-\pi}^{\pi} d\varphi\ \mathcal{J}_2 \lc 2 Z \sin \varphi \rc
\lL e^{2 i [n-1] \varphi} - e^{2 i [n+1] \varphi}  \rL =
2 \pi \bigg[ {\mathfrak{M}_1}_n - {\mathfrak{M}_2}_n +  {\mathfrak{P}_2}_n \bigg]
\label{eq_mod-neumann-solution}
}
$\forall\ n \geq 1$ which is in accordance with the series in Eq.~\eqref{eq_totW-circ-zprop-4} starting with $n=1$.
The additional definitions ${\mathfrak{M}_1}_n$, ${\mathfrak{M}_2}_n$ and ${\mathfrak{P}_2}_n$ are also listed in App.~\ref{app_definitions}. Finally, we obtain the following total probability:
\begin{framed}
\eqn{
W_{cs,\mathrm{coll}} = \frac{e^2    [2 \pi]   }{2^3  [2 q_{i,0}]  }
\sum_{n = 1}^{\infty} \int_{0}^{u_{n,\mathrm{coll}}} du_{cs}\ \hat{\mathfrak{J}}_{n,\mathrm{coll}}(Z_{n,\mathrm{coll}}(u_{cs}) ).
\label{eq_totW-circ-zprop-5}
}
\end{framed}
The non-trivial contribution has been abbreviated as follows:
\eqnsplit{
\hat{\mathfrak{J}}_{n,\mathrm{coll}}(Z_{n,\mathrm{coll}}) &\equiv 
\frac{1}{[1+u_{cs}]^2} \bigg[
- \frac{  \mathfrak{T}_{31,\mathrm{coll}} }{2}  \mathcal{J}_n^2
+ \frac{  \mathfrak{T}_{32,\mathrm{coll}} }{8} \lL \mathcal{J}_{n-1}^2 + \mathcal{J}_{n+1}^2  - 2 \mathcal{J}_n^2  \rL\\
&+ \frac{  \mathfrak{R}_{21,\mathrm{coll}} }{8}   \lL \mathcal{J}_{n-1}^2  - \mathcal{J}_{n+1}^2  \rL
+ \frac{  \mathfrak{R}_{21,\mathrm{coll}} }{ 8 } 
\big[  {\mathfrak{M}_1}_n - {\mathfrak{M}_2}_n  +  {\mathfrak{P}_2}_n   \big]
\bigg]
\label{eq_integrand-ncz}
}
where all functions 
$\mathcal{J}_n$, ${\mathfrak{M}_1}_n$, ${\mathfrak{M}_2}_n$ and ${\mathfrak{P}_2}_n$ 
have $Z_{n,\mathrm{coll}}$ as their argument.
Expression \eqref{eq_integrand-ncz} illustrates strongly that non-negligible extensions, in particular for high-energy photons, are modified by a set of standard Bessel and generalised hypergeometric functions\footnote{Definitions in App.~\ref{app_definitions}: Heaviside step functions are exactly highlighting this behaviour.}.

There is another reason
for studying a non-constrained solution with respect to the photon energies which has been aimed in 
the present study.
In several nonlinear QED scenarios it is common to probe the nonlinearity by an external pure electric plane-wave field provided, for instance, in some laser facilities. Assuming such a background field will immediately lead to the intensity parameter $\xi^2 =  e^2 E^2 [m \omega]^{-2}$ \cite{greiner}, where $E$ denotes the electric field amplitude and $\omega$ its frequency. For simplifying reasons allowing a perturbative treatment of the background field would therefore require $\xi \ll 1$. This limit can be either achieved for small electric field amplitudes or for sufficient large field frequencies. It becomes therefore evident that in latter case one has to take a more generalised result into account without constraining photon energies to the soft regime.

\subsection{Discussion on energetic contributions}

In this section we compare the unconstrained solution \eqref{eq_totW-circ-zprop-5}, derived in the previous section, with the in the literature well-known SPL approximation (App.~\ref{app_cs-spl}), considering the corresponding phase-space integrands, cf. Eqs.~\eqref{eq_integrand-ncz} and \eqref{eq_integrand-spl}, $\hat{\mathfrak{J}}_{n,\mathrm{coll}}(Z_{n,\mathrm{coll}})$ and $\hat{\mathfrak{J}}_{n,\mathrm{soft}}(Z_{n,\mathrm{soft}})$, normalised by $[m \xi]^{-2}$. Obviously, these terms contribute essentially to the total Compton scattering probability.
In case the condition \eqref{eq_ncz-condition} or \eqref{eq_ncz-condition2}, respectively, is fulfilled, the unconstrained formula is expected to become relevant for a proper description. According to this condition, we have to take all additive terms into account only if $\omega \gg m$, even in case of large parameters $\xi$. 
This limit indicates that for low-frequent fields the SPL solution would be sufficient, in particular in the environment of usually considered high intense external fields. However, in the energetic regime, i.e. field photon energies $\omega \gtrsim m$, this approximation does not hold. 
A first example is shown in Fig.~\ref{fig_ncz-1} below.

\begin{figure}[h]
\centering
\begin{minipage}{0.49\linewidth}
\centering
\subfigure[{red ($n=1$), blue ($n=2$), green ($n=3$) and dashed ($\mathrm{soft}$), solid ($\mathrm{coll}$)}]{\includegraphics[width=0.95\textwidth]{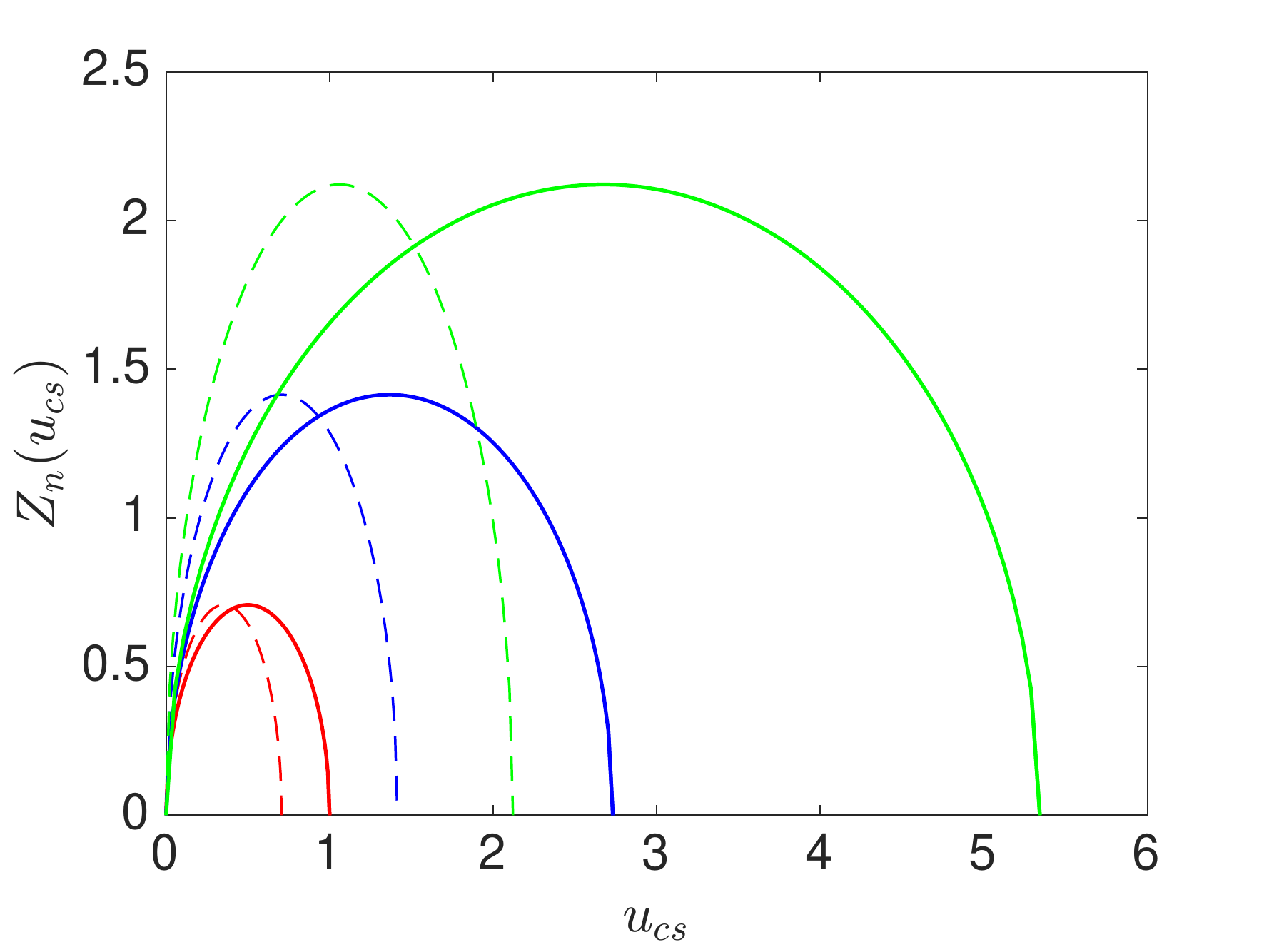}}
\end{minipage}
\begin{minipage}{0.49\linewidth}
\centering
\subfigure[{red ($n=1$), blue ($n=2$), green ($n=3$) and dotted envelope ($\mathrm{soft}$), solid envelope ($\mathrm{coll}$)}]{\includegraphics[width=0.95\textwidth]{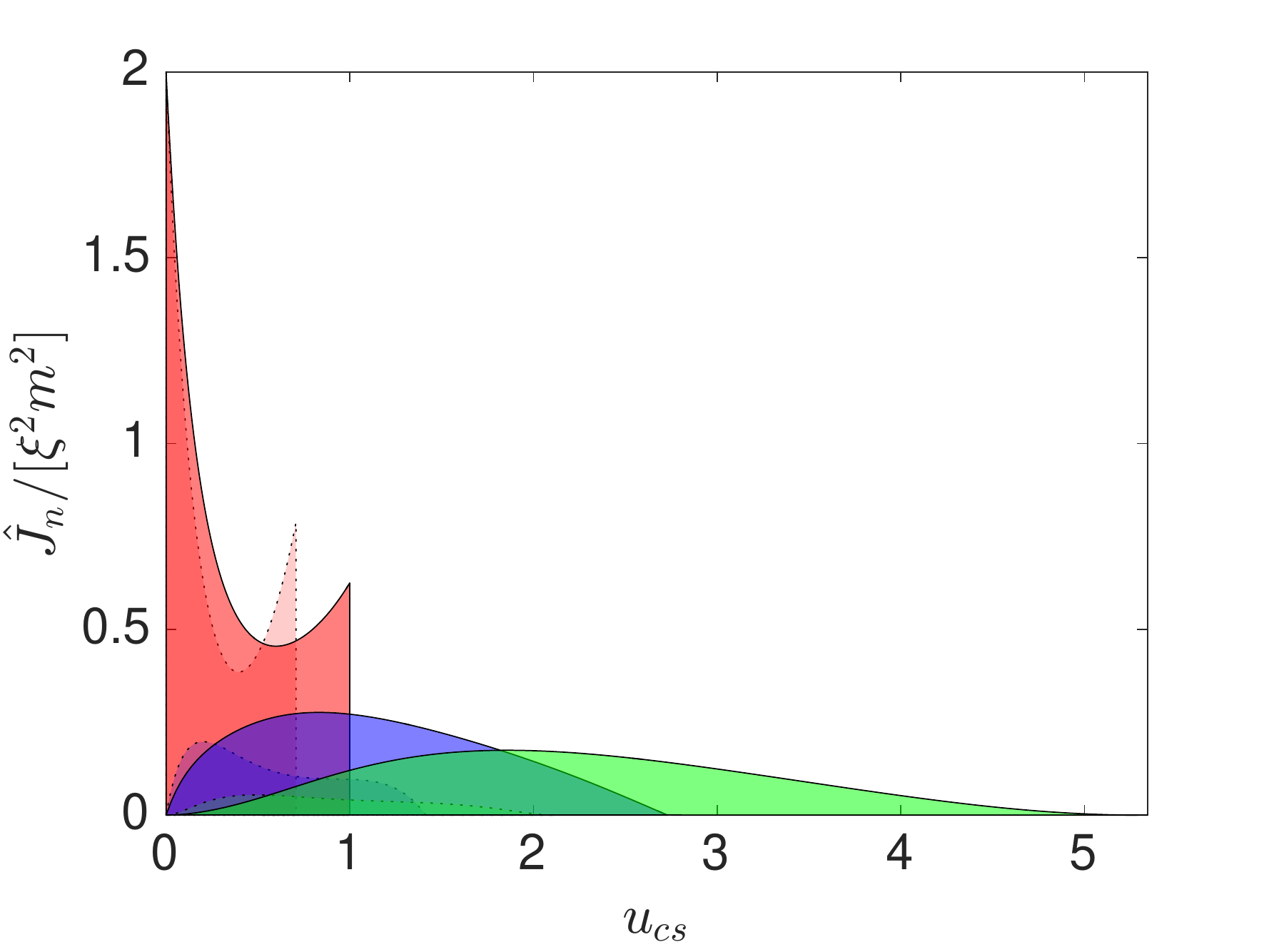}} 
\label{subfig_dressed_vertex}
\end{minipage}
\caption{Contributions to the total the Compton scattering probability for field parameters $\xi = 1.0$ and $\omega\ [m^{-1}]  = 0.5$: In the left panel the Bessel function arguments $Z_{n,\mathrm{coll}}(u_{cs}),Z_{n,\mathrm{soft}}(u_{cs})$ are plotted for the first three harmonics in the series. In the right panel the related integrands $\hat{\mathfrak{J}}_{n,\mathrm{coll}}(Z_{n,\mathrm{coll}})$ and $\hat{\mathfrak{J}}_{n,\mathrm{soft}}(Z_{n,\mathrm{soft}})$, normalised by $[m \xi]^{-2}$, and their integral areas are shown. The functions are plotted versus the integral variable $u_{cs}$ up to the corresponding Compton edges $u_{n,\mathrm{coll}}$ and $u_{n,\mathrm{soft}}$.}
\label{fig_ncz-1}
\end{figure}
 
\begin{figure}[h]
\centering
\begin{minipage}{0.49\linewidth}
\centering
\subfigure[ {Harmonic mode $n=2$: $\omega\ [m^{-1}] = \{1,1.2,1.4,1.6\}$ (red, blue, green, yellow) plotted for approaches $\mathrm{soft}$ (dotted envelope) and $\mathrm{coll}$ (solid envelope)}  ]{\includegraphics[width=0.95\textwidth]{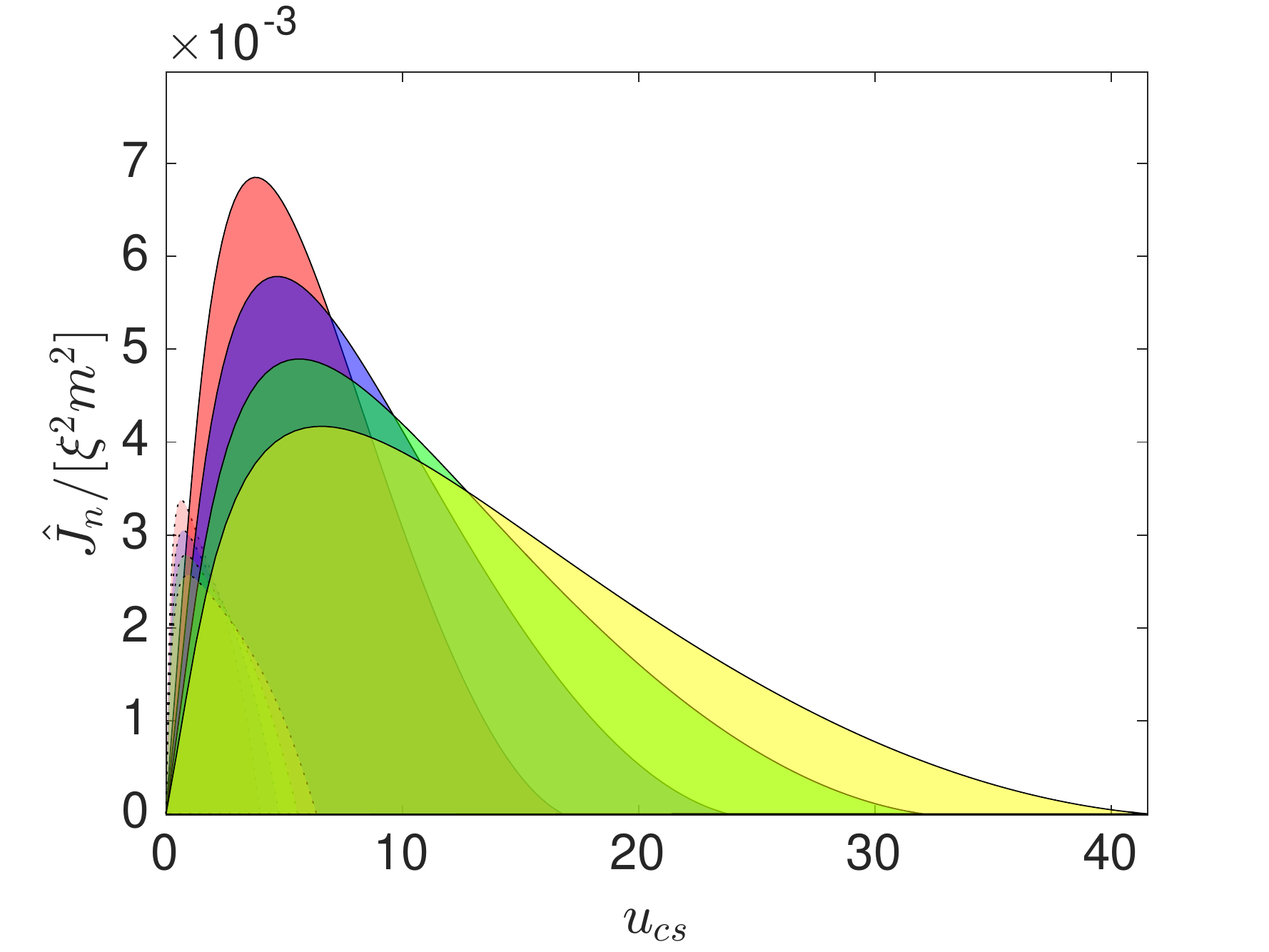}}
\end{minipage}
\begin{minipage}{0.49\linewidth}
\centering
\subfigure[ {Harmonic mode $n=3$: $\omega\ [m^{-1}] = \{1,1.2,1.4,1.6\}$ (red, blue, green, yellow) plotted for approaches $\mathrm{soft}$ (dotted envelope) and $\mathrm{coll}$ (solid envelope)} ]{\includegraphics[width=0.95\textwidth]{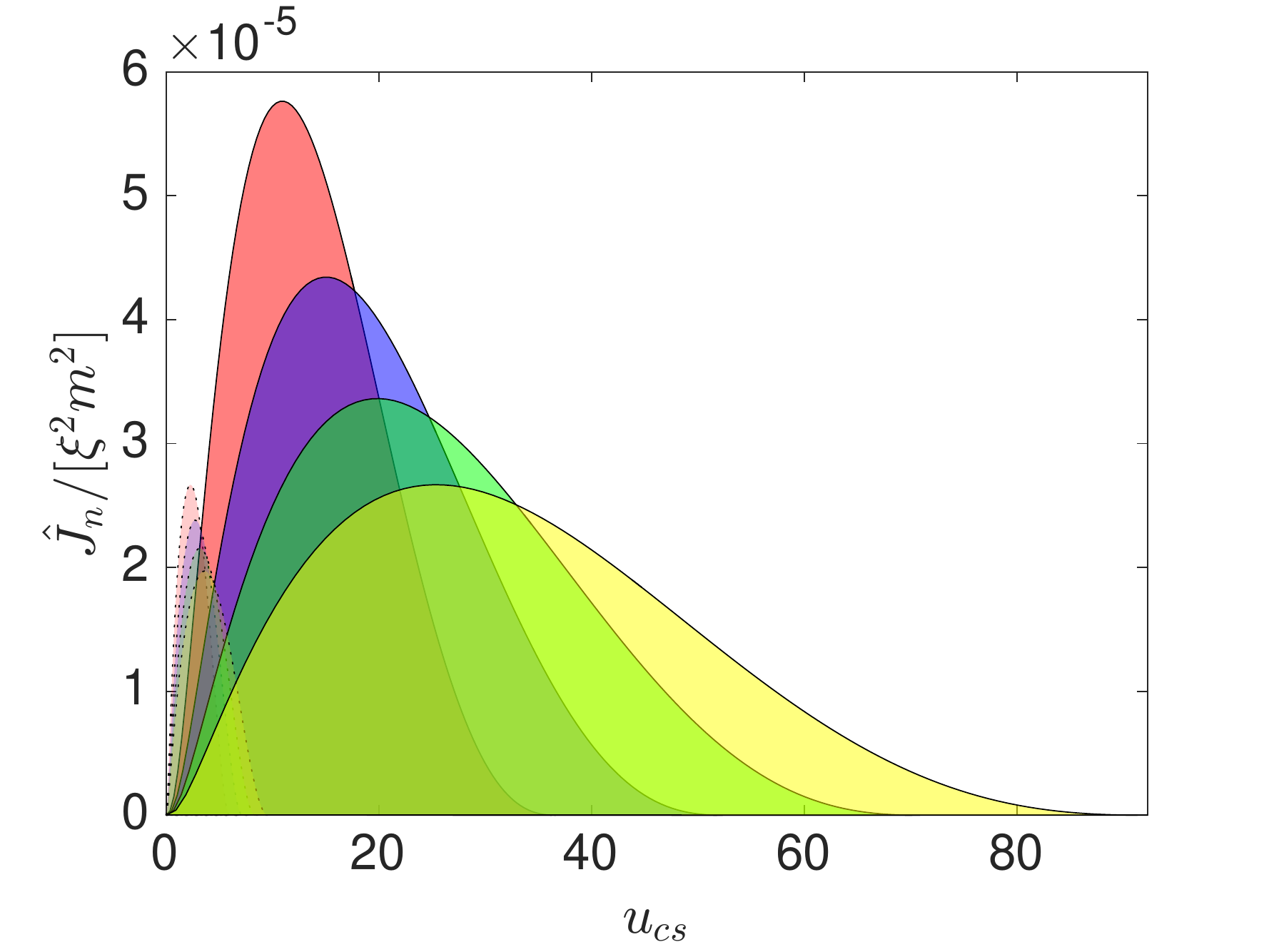}} 
\label{subfig_dressed_vertex}
\end{minipage}
\caption{Contributions to the total Compton scattering probability for fixed harmonic modes with $\xi = 0.1$: The integrands $\hat{\mathfrak{J}}_{n,\mathrm{coll}}(Z_{n,\mathrm{coll}})$  and $\hat{\mathfrak{J}}_{n,\mathrm{soft}}(Z_{n,\mathrm{soft}})$, normalised by $[m \xi]^{-2}$, are plotted for different photon energies versus $u_{cs}$ up to the corresponding Compton edges $u_{n,\mathrm{coll}}$ and $u_{n,\mathrm{soft}}$, respectively.}
\label{fig_ncz-2}
\end{figure}

We set the parameters, just for illustrative reasons, $\xi = 1.0$ and $\omega = 0.5 m$. On the left hand-side of Fig. \ref{fig_ncz-1} the corresponding function argument $Z_{n,\mathrm{coll}}(u_{cs})$ ($Z_{n,\mathrm{soft}}(u_{cs})$), cf. Eqs.~\eqref{eq_Z-ncz-final} and \eqref{eq_Z-spl-final-gen}, has been plotted versus $u_{cs}$ for the first three harmonic modes $n=1,2,3$. Note that the upper limit for $u_{cs}$ is exactly the Compton edge $u_{n,\mathrm{coll}}$ ($u_{n,\mathrm{soft}}$). 
Furthermore, Compton edges become more shifted with increasing $n$, whereas this effect is stronger for the unconstrained case according to \eqref{eq_Z-ncz-final}. It is remarkable that condition \eqref{eq_ncz-condition2} is already fulfilled for $n=1$, i.e. $u_{n,\mathrm{soft}} \sim 1$. Therefore, one expects a notable difference between  $\hat{\mathfrak{J}}_{n,\mathrm{coll}}$ and $\hat{\mathfrak{J}}_{n,\mathrm{soft}}$. This discrepancy is shown in Fig.~\ref{fig_ncz-1} (b). 
One observes a clear relevance of the unconstrained result, in particular becoming more distinct for higher $n$. 
This change is basically caused by two main reasons: due to larger Compton edges $u_{n,\mathrm{coll}}$, the integral area is enlarged. On the other hand, the envelope shape in the unconstrained case is modified due to additional terms, cf. Eq.~\eqref{eq_integrand-ncz}, which contribute in a crucial way to the complete integral area. One also has to note that in both cases the considered patterns --- so the complete integral value --- become narrowed. This clarifies for $n \rightarrow \infty$ the expected convergence of the series in Eqs.~\eqref{eq_totW-circ-zprop-5} and \eqref{eq_totW-spl-5}, respectively.
The impact of higher photon energies is shown in Fig.~\ref{fig_ncz-2}. 

Here we have $\xi = 0.1$ and fixed harmonic modes $n=2$ (a) and $n=3$ (b),  varying photon energies $\omega\ [m^{-1}] = \{1,1.2,1.4,1.6\} $. The leading term for $n=1$ is omitted but behaves in a similar way.
Due to the large discrepancies between the Compton edges $u_{n,\mathrm{soft}}$ and $u_{n,\mathrm{coll}}$, the unconstrained integrands exceed the SPL approach already at relative small $u_{cs}$ such that the SPL does not fit well to the exact total scattering probability. Furthermore, one recognizes that integral areas become drastically enlarged by increasing the photon energy. From this point of view, it might be interesting to study the effectiveness considering either large intensities or high energies. 
Investigations in this direction we plan to address in a forthcoming paper.
However, in total, latter features underline substantially the relevance of the generalised formula presented in \eqref{eq_totW-circ-zprop-5}.

\begin{figure}[h]
\centering
\begin{minipage}{0.32\linewidth}
\centering
\subfigure[$n=1$]{\includegraphics[width=1.0\textwidth]{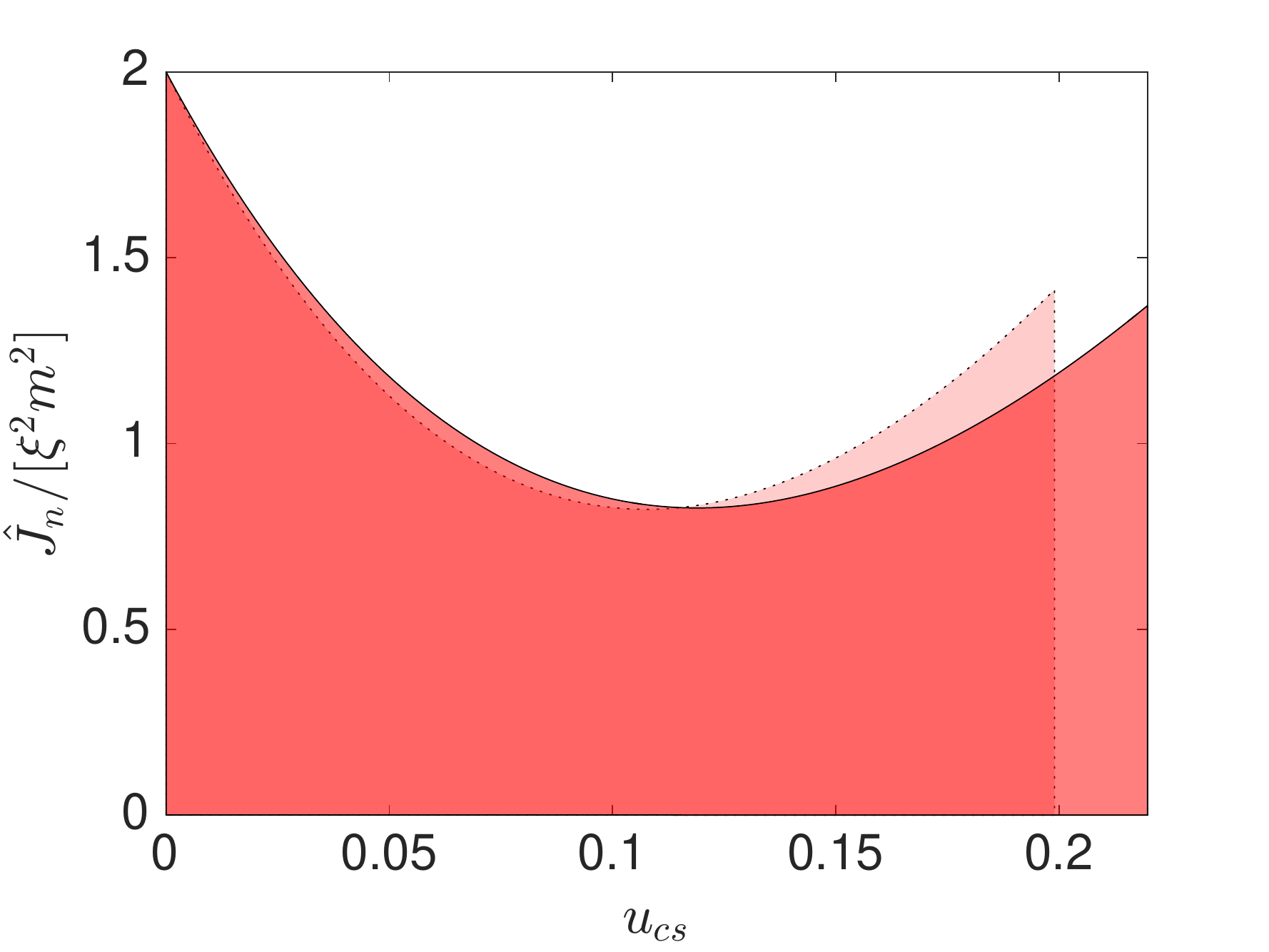}}
\end{minipage}
\begin{minipage}{0.32\linewidth}
\centering
\subfigure[$n=2$]{\includegraphics[width=1.0\textwidth]{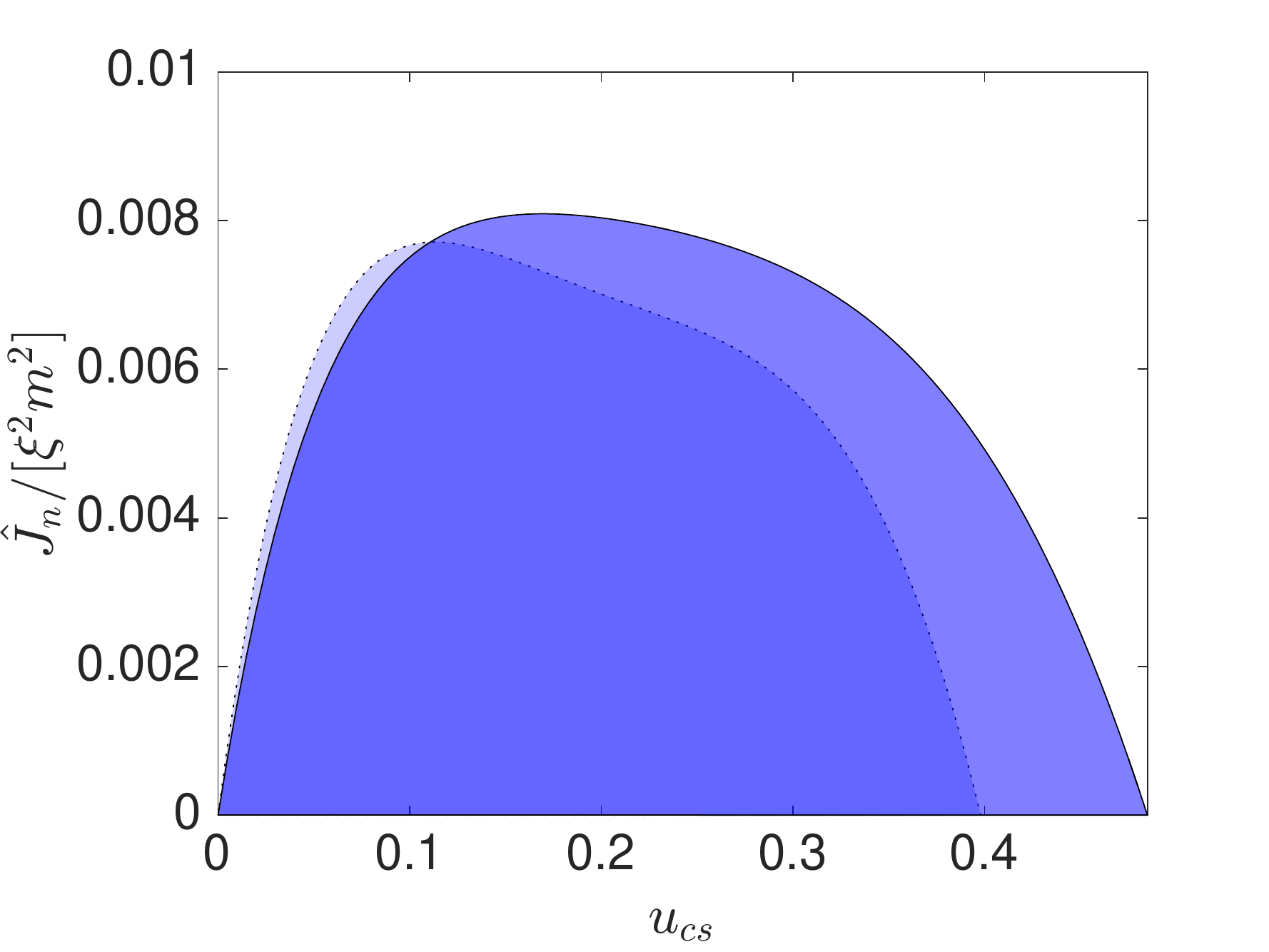}}
\end{minipage}
\begin{minipage}{0.32\linewidth}
\centering
\subfigure[$n=3$]{\includegraphics[width=1.0\textwidth]{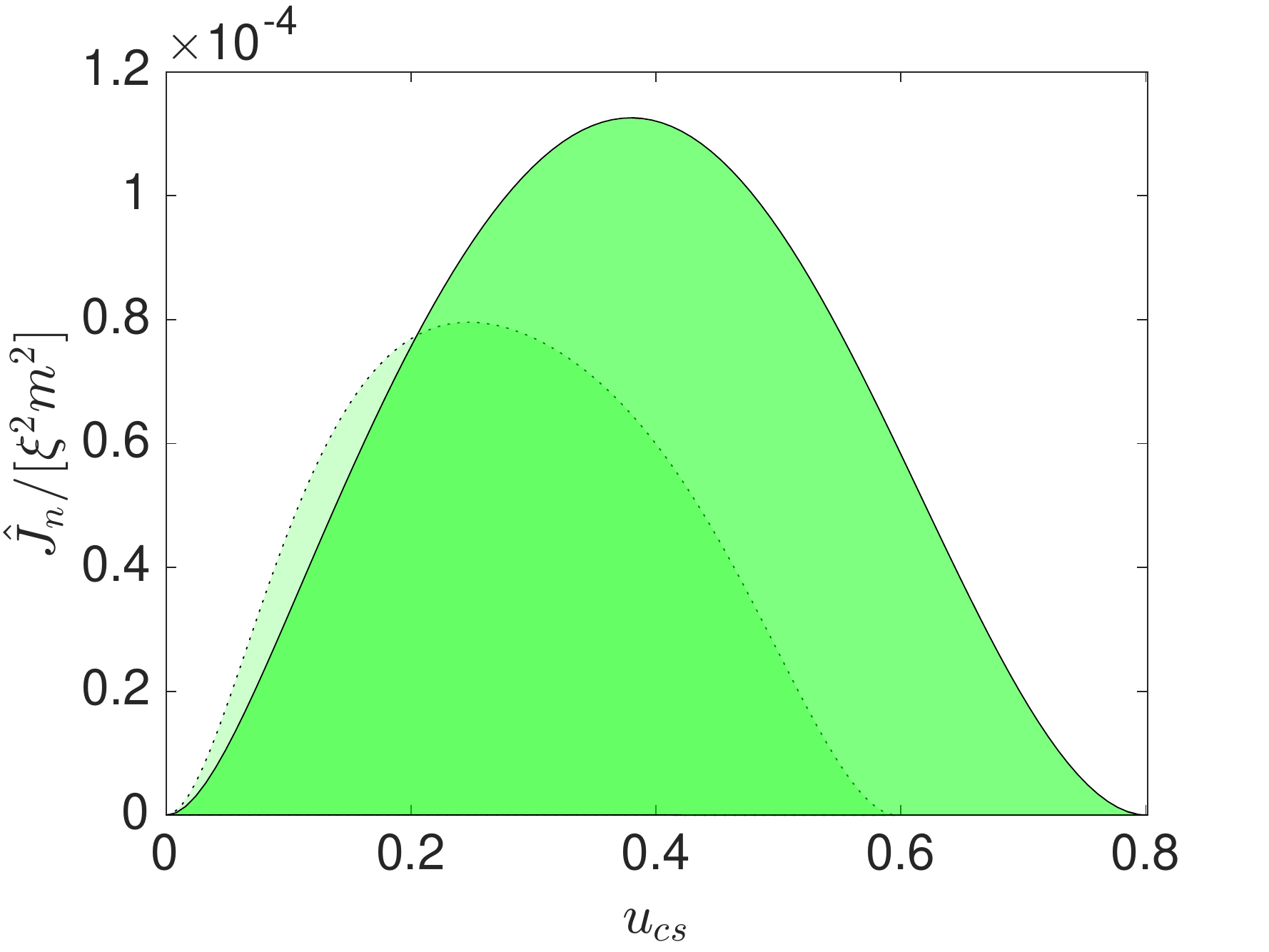}} 
\label{subfig_dressed_vertex}
\end{minipage}
\caption{Contributions to the total Compton scattering probability for parameters $\xi = 0.1$ and $\omega\ [m^{-1}] = 0.1$: Integrands $\hat{\mathfrak{J}}_{n,\mathrm{coll}}(Z_{n,\mathrm{coll}})$ (solid envelope) and $\hat{\mathfrak{J}}_{n,\mathrm{soft}}(Z_{n,\mathrm{soft}})$ (dotted envelope), normalised by $[m \xi]^{-2}$, are shown. The functions are plotted versus $u_{cs}$ up to the corresponding Compton edges $u_{n,\mathrm{coll}}$ and $u_{n,\mathrm{soft}}$.}
\label{fig_ncz-3}
\end{figure}

In a final step let us discuss the case of only small field parameters such that according to condition \eqref{eq_ncz-condition} the discrepancies between the unconstrained formula \eqref{eq_totW-circ-zprop-5} and the SPL approximation \eqref{eq_totW-spl-5} are expected to be small, particularly for the leading term in the series, compared with the cases before. 
Therefore, we choose $\xi = 0.1$ and $\omega = 0.1 m$ and show the corresponding integrands  \eqref{eq_integrand-ncz}, \eqref{eq_integrand-spl} and integral areas, respectively, for the first three harmonics $n=1,2,3$ in Fig.~\ref{fig_ncz-3}. Apart from the fact that next to leading terms, i.e. $n \geq 2$, achieve very small integrand values, cf. \ref{fig_ncz-3} (b) and (c), one realizes in addition that integrand values for latter cases are drastically reduced comparing with the case $n=1$ shown in Fig.~\ref{fig_ncz-3} (a). 
Due to the smallness of the Compton edges $u_{n,\mathrm{coll}}$ and $u_{n,\mathrm{soft}}$, respectively, we determine for $n=1$ the accordance of both integrands for $u_{cs} \lesssim 0.1$, which is in good agreement with condition \eqref{eq_ncz-condition2}. 
With increasing $n$ the Compton edges drift apart. 
Consequently, discrepancies between both formulae become more noticeable. Strictly speaking, for higher harmonics one needs to take the unconstrained formula into account, even if the contributions become less dominant. 

\section{Pair production in generalised plane-wave background fields}
\label{sec_pp}

\subsection{Differential pair production probability}
\begin{figure}[h]
\centering
\begin{minipage}{0.45\linewidth}
\centering
\subfigure[``Dressed'' fermionic external lines and free vertex.]{\includegraphics[width=0.8\textwidth]{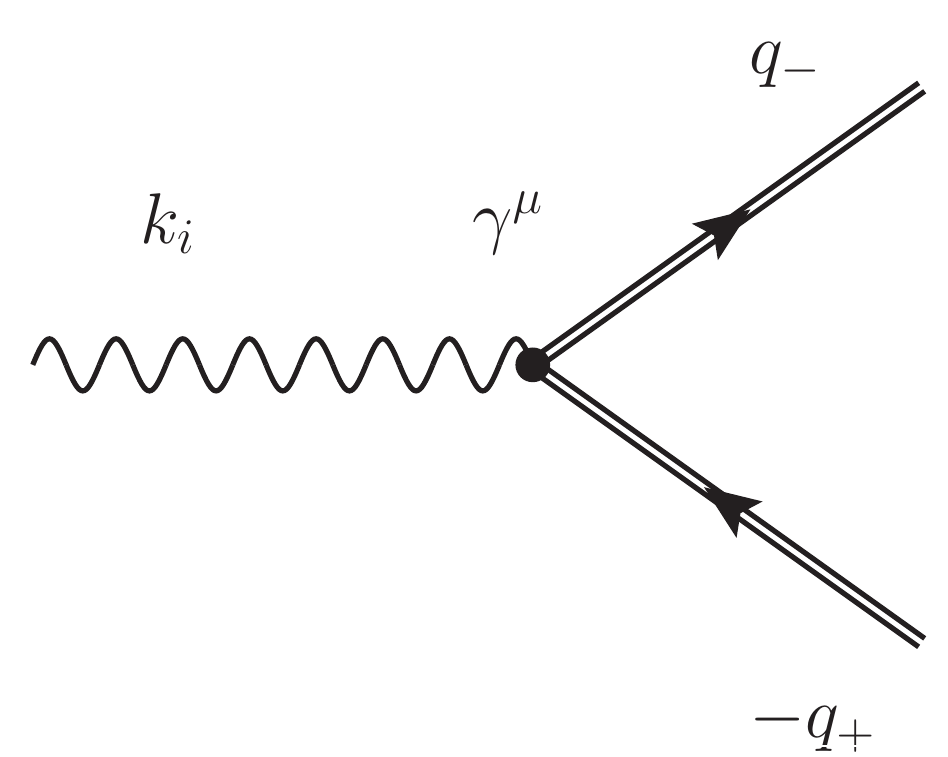}}
\end{minipage}
\begin{minipage}{0.45\linewidth}
\centering
\subfigure[Usual free fermionic external lines and ``dressed'' vertex.]{\includegraphics[width=0.8\textwidth]{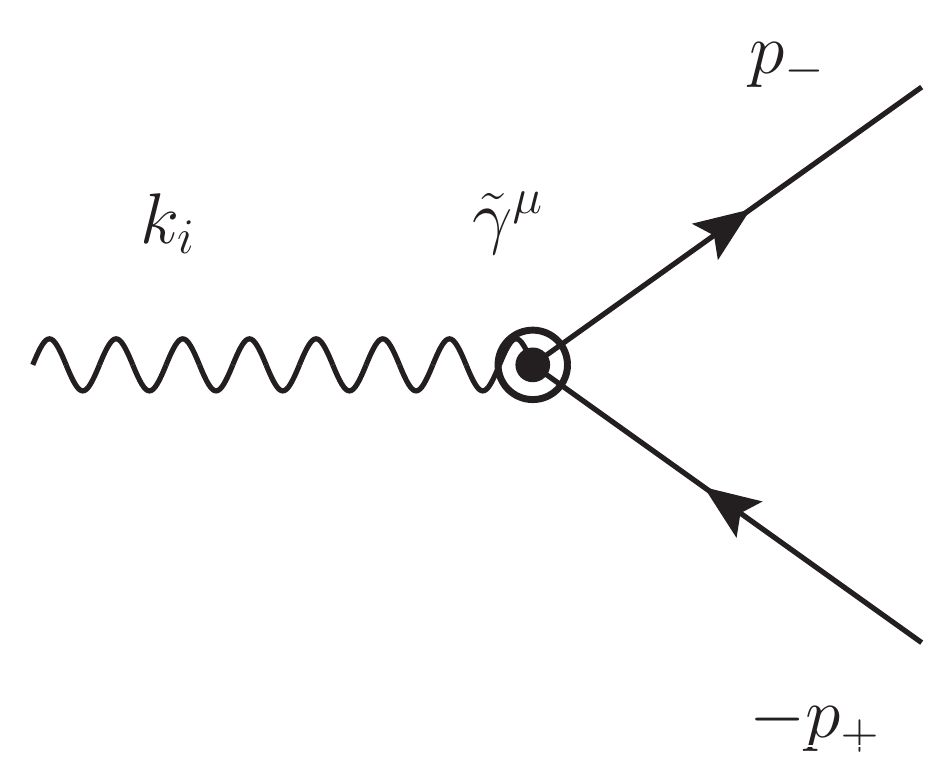}} 
\label{subfig_dressed_vertex}
\end{minipage}
\caption{Alternative diagrammatic illustrations for nonlinear Breit-Wheeler pair production in external background field. The modified ``dressed'' vertex $\tilde{\gamma}^\mu$ is provided with an additional cycle. Double lines are indicating the exact Volkov external lines for out-going electron and positron with momenta $q_+^\mu$ and $q_-^\mu$, respectively. The wavy line is representing the in-coming photon with momentum $k_i^\mu$. Note the negative momentum $-q_+^\mu$ for the anti-particle being ``time reversed''.}
\label{fig_pairp}
\end{figure}

In this section we will address another elementary QED process, namely the case of photon induced electron-positron pair production, often referred as the nonlinear Breit-Wheeler process \cite{breit-wheeler}. The mutual absorption of two photons was first discussed in Ref.~\cite{reiss} and is shown in Fig.~\ref{fig_pairp}. It is obtained from Fig.~\ref{fig_compton} via crossing symmetry, where $n$ external field quanta collide with a high-energy photon. 

Momentum conservation for the different harmonic modes reads as $s_n k^\mu+ k_i^\mu = q_-^\mu + q_+^\mu$, where $k_i^\mu$, $q_-^\mu$ and $q_+^\mu$ denote the momenta for the incoming photon, outgoing electron and outgoing positron, respectively, as previously defined in Eq.~\eqref{eq_defq_2}. Note, however, that unlike Compton scattering, strong field induced pair production is a threshold process which requires a minimum energy input or, equivalently, a minimum number of external field quanta discussed below, which becomes in SFQED $\frac{2 m^2}{k \cdot k_i} \rightarrow \frac{2 m_{\star}^2}{k \cdot k_i}$ \cite{bamber}. Following closely Sec.~\ref{subsec_genSmatrix}, we start with the S-matrix element
\eqn{
\mathcal{M}_{\pm} = \frac{-ie}{\sqrt{2 \omega_i}} \int d^4x\ \exp(i k_i \cdot x) \overline{\Psi^V_{ q_- }} \slashed{\varepsilon}_i \Psi^V_{ q_+ } ,
\label{eq_mfi-PP}
}
where ${\varepsilon_i}^\mu$ is the polarisation vector for the initial photon.
$\Psi^V_{ q_- }$ and $\Psi^V_{ q_+ }$ are external field dressed Volkov state representations for electron and positron, respectively. Performing the spatial integration after Fourier decomposing the modified vertex function, as carried out in Sec.~\ref{sec_scattering-gen}, leads to 
\eqnsplit{
\mathcal{M}_{\pm} &= -ie \frac{1}{\sqrt{ 2 \omega_i }}  \frac{1}{\sqrt{ 2 q_{+,0} }}  \frac{1}{\sqrt{ 2 q_{-,0} }} \frac{1}{2 \pi L} (2 \pi)^4  \sum_{n = - \infty}^{\infty}  \int_{- \pi L}^{ \pi L} d\varphi\\
&\times \delta^{(4)}\lc k_i  + s_n k - q_+  - q_- \rc
\exp \lc i s_n\varphi \rc
\overline u_{p_-}  \tilde{\gamma}^\mu(q_+,q_-,\varphi) {\varepsilon_i}_\mu  v_{p_+}
\label{eq_Mfi-PP}
}
with the modified vertex function
\eqn{
\tilde{\gamma}^\mu(q_+,q_-,\varphi) \equiv  \exp \lc i  \mathcal{P}_{q_-}(\varphi) \rc \Pi^\mu_{\varphi,pp} ( q_+ , q_-)  \exp \lc - i \mathcal{P}_{q_+}(\varphi)  \rc,
}
whereas $u_{p_-}$ ($v_{p_+}$) denotes the free Dirac bispinor for the electron (positron).
Note that the matrix resulting from the Fourier expansion, reads in case of pair production as:
\eqn{
\Pi^\mu_{\varphi,pp} ( q_+ , q_- ) \equiv \lL 1 + \frac{e  \slashed \gA(\varphi) \cdot \slashed k }{2 k \cdot q_- } \rL  \gamma^\mu  \lL 1 - \frac{e \slashed k \cdot \slashed  \gA(\varphi) }{2 k \cdot q_+ } \rL.
\label{eq_pi-PP}
}
Similarly to Eq.~\eqref{eq_dW-gen}, we have to compute again the square of $\mathcal{M}_{\pm}$ and sum over all final particle spins $s_-, s_+$ and over the initial photon polarisation $\varepsilon_i$. Therefore, one has to calculate the sum $\sum_{\varepsilon_i,s_-,s_+}  \lL \overline u_{p_-}  \mathrm{M}_{\varphi,pp}  v_{p_+} \rL \lL \overline u_{p_-}  \mathrm{M}_{\vartheta,pp}  v_{p_+} \rL^*$ with 
$\mathrm{M}_{\varphi,pp} \equiv \Pi^\mu_{\varphi,pp} ( q_+ , q_-)  {\varepsilon_i}_\mu$ and $\mathrm{M}_{\vartheta,pp} \equiv \Pi^\nu_{\vartheta,pp}  ( q_+ , q_-)  {\varepsilon_i}_\nu$. 
This leads to the trace
$\mathfrak{S}_{pp} \equiv \text{Tr} \lc [\slashed p_- + m] \Pi_{\varphi,pp} [\slashed p_+ - m] \gamma^0 \Pi^\dagger_{\vartheta,pp} \gamma^0  \rc$ for which we obtain by imposing the Lorenz gauge:
\eqnsplit{
\mathfrak{S}_{pp} = &- 16 m^2 - 8 p_- \cdot p_+
- 4 e \lL  \gA_\varphi + \gA_\vartheta  \rL \cdot \lL q_- \lL \frac{k \cdot q_+}{k \cdot q_-} + 1  \rL - q_+ \lL \frac{k \cdot q_-}{k \cdot q_+} + 1  \rL  \rL\\
&+ 4 e^2 \lL \gA_\vartheta^2 + \gA_\varphi^2  +  \gA_\varphi \cdot \gA_\vartheta  \lL \frac{k \cdot q_-}{k \cdot q_+} + \frac{k \cdot q_+}{k \cdot q_-} \rL   \rL,
\label{eq_tr-result-PP}
}
Rewriting $p_- \cdot p_+$, see Eq.~\eqref{eq_defq}, in terms of final effective momenta, we obtain,
similarly to Eq.~\eqref{eq_fin-dW}, the differential pair production probability:
\eqnsplit{
&dW_{pp} = e^2 [2 \pi]^4 \frac{1}{ 2 \omega_i } \frac{1}{ [2 \pi L]^2}   \frac{d^3\pmb{q}_-}{ 2 q_{-,0} } \frac{d^3\pmb{q}_+}{ 2 q_{+,0} } \sum_{n = - \infty}^{\infty} \int_{- \pi L}^{ \pi L} d\varphi \int_{- \pi L}^{ \pi L} d\vartheta\
\delta^{(4)}\lc k_i  + s_n k - q_+  - q_- \rc \times\\
& \bigg\{ 
- 16 m^2 - 8 q_- \cdot q_+  - 4 e^2 \mathfrak{a}^2 \lL \frac{k \cdot q_- }{k \cdot q_+ }  + \frac{k \cdot q_+ }{k \cdot q_- } \rL
+ 4 e^2 \lL \gA_\vartheta^2 + \gA_\varphi^2 +  \gA_\varphi \cdot \gA_\vartheta  \lL \frac{k \cdot q_-}{k \cdot q_+} + \frac{k \cdot q_+}{k \cdot q_-} \rL   \rL\\
&- 4 e \lL  \gA_\varphi + \gA_\vartheta  \rL \cdot \lL q_- \lL \frac{k \cdot q_+}{k \cdot q_-} + 1  \rL - q_+ \lL \frac{k \cdot q_-}{k \cdot q_+} + 1  \rL  \rL
\bigg\}
\exp \lc i \lL \frac{q_- \cdot \mathcal{B}_{(\varphi,\vartheta)} }{k \cdot q_-}  - \frac{q_+ \cdot \mathcal{B}_{(\varphi,\vartheta)} }{k \cdot q_+} \rL  \rc\\
&\times \exp \lc i s_n [\varphi - \vartheta] \rc.
\label{eq_fin-dW-PP}
}

\subsection{Phase-space in photon-photon collision}
\label{subsec_phocs_pp}
We choose again the external field photon momentum orientation in $\hat z$-direction, i.e.
$k^\mu = [\omega,0,0,\omega]$.
The invariant phase-space measure is given by
\eqn{
\int d \text{LIPS}\ \delta^{(4)}\lc k_i  + s_n k - q_+  - q_- \rc = \int \frac{d^3 \pmb{q}_-}{2 q_{-,0} }  \int \frac{d^3 \pmb{q}_+}{2 q_{+,0} }  \delta^{(4)}\lc k_i  + s_n k - q_+  - q_- \rc.
}
As in the case before, we define 
\eqn{
u_{pp} \equiv \frac{k \cdot q_+}{k \cdot q_-},
\label{eq_invariant-u-PP}
}
and we construct a mode-dependent $\text{CMS}_{n,pp}$ frame characterised by
\eqn{
s_n \pmb{k} + \pmb{k}_i = \pmb{q}_- + \pmb{q}_+ = 0.
\label{eq_cms-PP}
}
Regarding the interpretation of such a ``perturbative'' CMS frame, we refer to previous discussion in Sec.~\ref{subsec_phocs}.
Condition \eqref{eq_cms-PP} implies directly $|\pmb{q}_+|  = |\pmb{q}_-|$. For the total energy of the system we conclude accordingly
\eqn{
\epsilon_{n,pp} \equiv s_n \omega + \omega_i = q_{-,0} + q_{+,0}. 
\label{eq_system-energy-PP}
}
Imposing latter $\text{CMS}_{n,pp}$ frame in case of pair production does not lead to constraints for the photon energy. However, one has to take into account that imposing $\text{CMS}_{n,pp}$ leads to an initial quantised photon with momentum $\pmb{k}_i = [0,0,-\omega_i]$ in $\hat z$-direction. Therefore, 
\eqn{
s_n \omega = \omega_i
}
has to be set.
Consequently, the choice for $\pmb{k}_i$ leads to
\eqnsplit{ 
&\gA \cdot k_i = \omega_i \gA_{,3},\\
&\mathcal{B}_{(\varphi,\vartheta)} \cdot k_i = \omega_i \mathcal{B}_{(\varphi,\vartheta),3},
\label{eq_gauge-extension-PP}
}
cf. Eq.~\eqref{eq_defBcal}. Furthermore, the effective electron and positron momenta are supposed to be on-shell with respect to the effective quasi-mass, i.e. $q_-^2 = q_+^2 = m_{\star}^2$. One should note that those relations force additionally the equality
\eqn{
q_{-,0} = q_{+,0} = \omega_i
} 
in the present POCS-$\text{CMS}_{n,pp}$ frame.
We split up the resulting delta function in the following way
$\delta^{(4)}\lc k_i  + s_n k - q_+  - q_- \rc \longrightarrow \delta \lc \epsilon_{n,pp} - q_{-,0} - q_{+,0}\rc \delta^{(3)} \lc \pmb{q}_- + \pmb{q}_+ \rc$
and parametrize the final effective electron momentum $\pmb{q}_-$ applying spherical coordinates.
Note that the absolute value for the final created electron momentum is completely fixed due to POCS-$\text{CMS}_{n,pp}$ with $|\pmb{q}_-| = \sqrt{\omega_i^2 - m_{\star}^2}$. This indicates that the integration over all possible final electron momenta requires the condition $\omega_i \geq m_{\star}$, leading in the POCS-$\text{CMS}_{n,pp}$ frame immediately to the previous mentioned minimum number of external field quanta $s_n \geq 2 m_{\star}^2 [k \cdot k_i]^{-1}$.
Following closely Sec.~\ref{subsec_phocs}, we proceed first with 
\eqn{
u_{pp} + 1 = \epsilon_{n,pp}  [ q_{-,0} - |\pmb{q}_-| \cos(\theta) ]^{-1}
}
and obtain the differential relation
\eqn{
d\cos(\theta) = \frac{\epsilon_{n,pp}}{|\pmb{q}_- |[1+u_{pp}]^2} du_{pp},
\label{eq_dcosTheta-PP}
}
respectively. 
The corresponding  upper and lower integration limit, in analogy to Eqs.~\eqref{eq_int-up-limit} and \eqref{eq_int-low-limit}, but with respect to the system energy from \eqref{eq_system-energy-PP}, reads as:
\begin{align}
%
\max\{u_{pp}\} &\longrightarrow u_{n,pp}^{max} \equiv  \frac{\epsilon_{n,pp}^2[1 + \sqrt{ 1 - 4 m_{\star}^2 / \epsilon_{n,pp}^2 }]     }{ 2m_{\star}^2 } - 1 \quad  [ \theta = 0 ], \label{eq_int-up-limit-PP}\\
%
%
\min\{u_{pp}\} &\longrightarrow  u_{n,pp}^{min} \equiv \frac{\epsilon_{n,pp}^2[1 - \sqrt{ 1 - 4 m_{\star}^2 / \epsilon_{n,pp}^2 }]     }{ 2m_{\star}^2 } - 1 \quad [ \theta = \pi ]. \label{eq_int-low-limit-PP}
\end{align}
Hence, we derive a reduced Lorentz-invariant phase-space measure 
\eqn{
d\widehat{\text{LIPS}}_{pp}
\equiv \int_{0}^{2 \pi} d\phi \int_{ u_{n,pp}^{min} }^{ u_{n,pp}^{max} } \frac{d u_{pp}}{2^3 [1+u_{pp}]^2 }
\label{eq_reduced-ps-pp}
}
similar to \eqref{eq_reduced-ps}.

\subsection{Total pair production probability}

Taking the minimum number of field quanta into account ($s_n > 0$), we apply finally \eqref{eq_reduced-ps-pp} to Eq.~\eqref{eq_fin-dW-PP} and obtain: 
\eqnsplit{
&W_{pp} =  \frac{  e^2  }{ 2 \omega_i   } \frac{1}{[2 \pi L]^2}    \sum_{n = 1}^{\infty} \int_{- \pi L}^{ \pi L} d\varphi \int_{- \pi L}^{ \pi L} d\vartheta  \exp \lc i s_n [\varphi - \vartheta] \rc \int d\widehat{\text{LIPS}}_{pp}\\
&\times \bigg\{ 
-16 m^2 
- 8 q_- \cdot [ s_n k + k_i ]  + 8 m_{\star}^2 
- 4 e^2 \mathfrak{a}^2 \lL \frac{1 + u_{pp}^2}{ u_{pp} } \rL\\
&- 4 e \lL \gA_\varphi + \gA_\vartheta \rL \cdot \lL q_- \lL \frac{ [ 1 + u_{pp} ]^2 }{ u_{pp} }  \rL   - k_i \lL \frac{1 + u_{pp}}{ u_{pp} } \rL  \rL
+ 4 e^2 \lL \gA^2_\varphi + \gA^2_\vartheta + \gA_\varphi \cdot \gA_\vartheta \lL  \frac{1 + u_{pp}^2}{ u_{pp} } \rL  \rL 
\bigg\}\\
&\times \exp \lc i \lL  
\lL \frac{ [ 1 + u_{pp} ]^2 }{ u_{pp} }  \rL \frac{q_-  \cdot \mathcal{B}_{(\varphi,\vartheta)} }{k \cdot k_i}   - \lL  \frac{1 + u_{pp}}{ u_{pp} } \rL \frac{k_i \cdot \mathcal{B}_{(\varphi,\vartheta)} }{ k \cdot k_i} 
\rL  \rc.
\label{eq_totW1-PP}
}
Latter expression shows that applying a variable substitution $\tilde{u} = \frac{[1+ u_{pp} ]^2}{4 u_{pp}}$, would be principally also beneficial for solving the integration over $u_{pp}$ or respectively $\tilde u$. Indeed the choice $\tilde{u} \equiv  \frac{[k \cdot k_i]^2}{ 4  k \cdot q_-   k \cdot q_+ }$ has been considered in most of the discussions in the literature for some specific field shape assumptions, as for instance circular polarised shapes in \eqref{eq_circ-field}. Defining, analogously as in the case before, some $\phi$-independent functions $\tilde{\mathfrak{B}}_j$ and $\tilde{\mathfrak{T}}_j$, see App.~\ref{app_definitions}, we obtain finally the following generalised total pair production probability for the case $\omega_i \geq m_{\star}$:
\begin{framed}
\eqnsplit{
&W_{pp} = \frac{e^2   [2\pi]  }{2^3    [2 \omega_i] [2 \pi L]^2} \sum_{n = 1}^{ \infty} \int_{-\pi L}^{\pi L} d\varphi \int_{-\pi L}^{ \pi L} d\vartheta \exp\lc i s_n [\varphi - \vartheta] \rc 
\int_{ u_{n,pp}^{min}  }^{ u_{n,pp}^{max} }  \frac{d u_{pp} }{[1+u_{pp}]^2} \\
&\times
\bigg\{
\tilde{\mathfrak{T}}_3 \mathcal{J}_0 \lc  \sqrt{  \tilde{\mathfrak{B}}_1^2 + \tilde{\mathfrak{B}}_2^2 } \rc  
- \frac{ i  \lL  \tilde{\mathfrak{B}}_2 \tilde{\mathfrak{T}}_2  +  \tilde{\mathfrak{B}}_1 \tilde{\mathfrak{T}}_1  \rL   }{ 2 } \lL \mathcal{J}_0 \lc  \sqrt{  \tilde{\mathfrak{B}}_1^2 
+ \tilde{\mathfrak{B}}_2^2  } \rc + \mathcal{J}_2 \lc  \sqrt{  \tilde{\mathfrak{B}}_1^2 + \tilde{\mathfrak{B}}_2^2  } \rc   \rL
\bigg\} \\
&\times \exp\lc -i \tilde{\mathfrak{B}}_3  \rc.
\label{eq_gen-fin-totW-PP}
}
\end{framed}
One should note that our result in \eqref{eq_gen-fin-totW-PP} extends the expressions discussed in the literature, cf. e.g. Ref.~\cite{greiner}, where usually only the first term is contained, multiplied with $\tilde{\mathfrak{T}}_3$ in case of the field from \eqref{eq_circ-field}. Such a simplified expression, for instance, can precisely be derived on basis of the general result \eqref{eq_gen-fin-totW-PP}, assuming initial photon energies of the order $\omega_i \sim m_{\star}$ in addition. Hence, final electron momenta become negligible small, in particular $|\pmb{q}_-|^2 \approx 0$, such that according to $\tilde{\mathfrak{B}}_j \tilde{\mathfrak{T}}_j  \approx 0$ for $j=1,2$, the extra terms from above are negligible.

\newpage
\section{Conclusion}
\label{sec_out}
We presented a comprehensive investigation of nonlinear lepton-photon interactions in external background fields.
The considered strong-field processes were Compton scattering and stimulated electron-positron pair production (nonlinear Breit-Wheeler process).
A semi-classical method based on coherent states of radiation allowed us to treat the
external background quasi-classically in the ordinary QED action. 
For the corresponding modified particle states we
applied the well-known Volkov representations where the background field has been assumed to be an arbitrary periodic plane-wave.
Constructing a special photon-oriented coordinate frame, we obtained for a large class of plane-wave background fields a closed formula for the total scattering probability considering S-matrix elements without constraining photon energies. Our result includes in its most general form a sum over all harmonics coming from the field quanta, a one-fold phase-space integration, plus a two-fold integral over background induced phases.
We obtained additive contributions extending known expressions in the literature beyond the usual soft terms.
We discussed in great detail the relevance of these extra terms
by applying our general formula to Compton scattering by an electron propagating in a laser-like background and compared our unconstrained phase-space integrand with the one in SPL. 
We showed that already the leading term in the soft limit is not sufficient to describe the exact total scattering probability for large energies.
Furthermore, we calculated that the generalised unconstrained solution is strongly required in particular for the highly nonlinear regime. In case of small field parameters those effects become already relevant in the second harmonics.
Applying crossing symmetry, we derived accordingly for the same class of external background fields a closed formula for the total pair production probability. Interacting photons have been considered energetically unconstrained such that 
new additive contributions, comparing with expressions from the literature, become relevant as in the case before.

\section{Acknowledgements}
I.A. would like to thank Carsten M{\"{u}}ller and Andreas Ringwald for their useful comments on the manuscript. This work has received the support of the Collaborative Research Center SFB 676 ``Particles, Strings, and the Early Universe'' of the DFG.

\newpage
\begin{appendix}

\section{Definitions, solutions and abbreviations}
\label{app_definitions}

%
\subsection*{Definitions of $\phi$-independent functions in POCS-$\text{CMS}_{n,cs}$}
\begin{align}
\mathfrak{B}_1 &\equiv  [1+u_{cs}] \omega_f \mathcal{B}_{(\varphi,\vartheta),1} \sin(\theta) [k \cdot q_i]^{-1} \label{def_B1}\\
\mathfrak{B}_2 &\equiv  [1+u_{cs}] \omega_f \mathcal{B}_{(\varphi,\vartheta),2} \sin(\theta) [k \cdot q_i]^{-1} \label{def_B2}\\
\mathfrak{B}_3 &\equiv  [1+u_{cs}] \omega_f \mathcal{B}_{(\varphi,\vartheta),3} \cos(\theta) [k \cdot q_i]^{-1}
+ q_i \cdot \mathcal{B}_{(\varphi,\vartheta)} [k \cdot q_i]^{-1} \label{def_B3}\\
\mathfrak{T}_1 &\equiv   4 e u_{cs} \omega_f [\mathcal{A}_{\varphi} + \mathcal{A}_{\vartheta}]_{,1} \sin(\theta) \label{def_T1}\\
\mathfrak{T}_2 &\equiv   4 e u_{cs} \omega_f [\mathcal{A}_{\varphi} + \mathcal{A}_{\vartheta}]_{,2} \sin(\theta) \label{def_T2}\\
\mathfrak{T}_3 &\equiv 16 m^2 - 8 q_{i,0} q_{f,0} + 8 \omega_f q_{i,3} \cos(\theta)
- 4 e u_{cs} q_i \cdot  [\mathcal{A}_{\varphi} + \mathcal{A}_{\vartheta}] [1+u_{cs}]^{-1} \notag\\
&- 4e^2 \lL \mathcal{A}_{\varphi}^2 + \mathcal{A}_{\vartheta}^2  - \mathcal{A}_{\varphi} \cdot \mathcal{A}_{\vartheta}  [1 + [1+u_{cs} ]^2][1+u_{cs} ]^{-1}  \rL  \notag\\
& + 4e u_{cs} \omega_f [\mathcal{A}_{\varphi} +  \mathcal{A}_{\vartheta}]_{,3} \cos(\theta)   - 4 e^2 \mathfrak{a}^2 [1+[1+ u_{cs} ]^2] [1+ u_{cs} ]^{-1}
\label{def_T3}
\end{align}

\subsection*{Definitions of $\phi$-independent functions in POCS-$\text{CMS}_{n,pp}$}

\begin{align}
\tilde{\mathfrak{B}}_1 &\equiv  [1+u_{pp}]^2 u_{pp}^{-1} |\pmb{q}_-| \mathcal{B}_{(\varphi,\vartheta),1} \sin(\theta) [k \cdot k_i]^{-1}\\
\tilde{\mathfrak{B}}_2 &\equiv  [1+u_{pp}]^2 u_{pp}^{-1} |\pmb{q}_-| \mathcal{B}_{(\varphi,\vartheta),2} \sin(\theta) [k \cdot k_i]^{-1}\\
\tilde{\mathfrak{B}}_3 &\equiv  [1 + u_{pp}] u_{pp}^{-1}  k_i  \cdot \mathcal{B}_{(\varphi,\vartheta)} [k \cdot k_i]^{-1}
- [1+u_{pp}]^2 u_{pp}^{-1} q_- \cdot  \mathcal{B}_{(\varphi,\vartheta)} [k \cdot k_i]^{-1}\\
\tilde{\mathfrak{T}}_1 &\equiv  4 e  [1+u_{pp}]^2 u_{pp}^{-1}  |\pmb{q}_-| [\mathcal{A}_{\varphi} + \mathcal{A}_{\vartheta}]_{,1} \sin(\theta)\\
\tilde{\mathfrak{T}}_2 &\equiv  4 e  [1+u_{pp}]^2 u_{pp}^{-1}  |\pmb{q}_-| [\mathcal{A}_{\varphi} + \mathcal{A}_{\vartheta}]_{,2} \sin(\theta)\\
\tilde{\mathfrak{T}}_3 &\equiv - 16 m^2 + 8 m_{\star}^2
 - 16 \omega_i^2  - 4 e^2 \mathfrak{a}^2 [1+u_{pp}^2] u_{pp}^{-1} \notag\\
&+ 4e^2 \lL \mathcal{A}_{\varphi}^2 + \mathcal{A}_{\vartheta}^2  + \mathcal{A}_{\varphi} \cdot \mathcal{A}_{\vartheta}  [1+u_{pp}^2] u_{pp}^{-1}  \rL 
+ 4 e k_i \cdot \lL  \mathcal{A}_{\varphi} + \mathcal{A}_{\vartheta}  \rL  [1 + u_{pp}] u_{pp}^{-1}  \notag\\ 
&+ 4e [1+u_{pp}]^2 u_{pp}^{-1}  |\pmb{q}_-| [\mathcal{A}_{\varphi} + \mathcal{A}_{\vartheta}]_{,3} \cos(\theta) 
\label{eq_def-abbrevs-PP}
\end{align}

\subsection*{Solution of modified Neumann's integral identity}

The solution for the modified Neumann identity reads as
\eqnsplit{
&\int_{-\pi}^{ \pi} d\varphi\ \mathcal{J}_2 \lc 2 Z \sin\varphi  \rc \exp(2 i n \varphi) = 2 \pi \lL  \Theta\lc \frac{1}{2} -  |n| \rc \mathcal{J}_{n+1}^2(Z) - \Theta\lc n - \frac{1}{2} \rc \mathfrak{F}_n(Z) \rL,
} 
valid $\forall\ n \geq 0\ \forall\ Z \geq 0$ where 
\eqnsplit{
\mathfrak{F}_n(Z) &\equiv 2^{-4-2 n} Z^{2 n} 
\bigg[  
-\frac{16 \pFq{1}{2}{-\frac{1}{2} + n}{1 + n,-1 + 2 n}{-Z^2}  }{ \Gamma(n) \Gamma(1+n) }\\  
&+ \frac{8 Z^2 \pFq{1}{2}{ \frac{1}{2} + n}{2 + n, 1 + 2 n}{-Z^2}  }{ \Gamma(1+n) \Gamma(2+n) }\\
&- \frac{Z^4 \pFq{1}{2}{ \frac{3}{2} + n}{3 + n, 3 + 2 n}{-Z^2}  }{ \Gamma(2+n) \Gamma(3+n) }\\
&+ \frac{n 16 \pFq{2}{3}{ -\frac{1}{2} + n, 1+n }{n, 2+n, -1 + 2 n}{-Z^2}  }{ \Gamma(n) \Gamma(2+n) }\\
&- \frac{8 [1+n] Z^2 \pFq{2}{3}{ \frac{1}{2} + n, 2+n }{1+n, 3+n, 1 + 2 n}{-Z^2}  }{ \Gamma(1+n) \Gamma(3+n) }\\
&+ \frac{ [2+n] Z^4 \pFq{2}{3}{ \frac{3}{2} + n, 3+n }{2+n, 4+n, 3 + 2 n}{-Z^2}  }{ \Gamma(2+n) \Gamma(4+n) }
\bigg]
}
has been defined. Here, ${}_p \mathcal{F}_q$ denotes the generalised hypergeometric function of order $p, q$ and is defined as:
\eqn{
{}_p \mathcal{F}_q(\{a_1,\ldots,a_p\};\{b_1,\ldots,b_q\};c) \equiv \sum_{k = 0}^{\infty} \frac{(a_1)_k (a_2)_k \ldots (a_p)_k}{ (b_1)_k (b_2)_k \ldots (b_q)_k } \lL \frac{c^k}{k!} \rL.
\label{eq_pFq}
}
$(a)_k$ and $(b)_k$ denote the Pochhammer symbol
\eqn{
(a)_k \equiv \frac{\Gamma(a + k)}{\Gamma(a)}
\label{eq_pochhammer}
}
in terms of the usual Gamma function $\Gamma$, which is defined for complex numbers with a positive real part via a convergent improper integral as follows:
\eqn{
\Gamma(t) \equiv \int_0^\infty x^{t-1} \exp(-x)\ dx.
\label{eq_def-GammaFunction}
}
Note, if $t \in\ \mathbb{N}_+$, the Gamma function can be related to the factorial by
\eqn{
\Gamma(t) = (t-1)!.
\label{eq_def-GammaFunction2}
}

\subsection*{Functions resulting from modified Neumann's integral identity}

\begin{align}
{\mathfrak{M}_1}_n(Z)   &\equiv  \Theta\lc \frac{1}{2} -  |n-1| \rc \mathcal{J}_{n}^2(Z)\\
{\mathfrak{M}_2}_n(Z)   &\equiv  \Theta\lc [n-1] - \frac{1}{2} \rc \mathfrak{F}_{n-1}(Z)\\
{\mathfrak{P}_2}_n(Z)   &\equiv  \Theta\lc [n+1] - \frac{1}{2} \rc \mathfrak{F}_{n+1}(Z)
\end{align}

\section{Soft-photon limit for Compton scattering}
\label{app_cs-spl}

The SPL approach for Compton scattering is considered, for instance in Refs.~\cite{landau, ritus}, assuming the same circularly polarised field  from Eq.~\eqref{eq_circ-field}  we have applied before.
In this Appendix we illustrate the direct reduction to the total probability in SPL starting from our general result \eqref{eq_totW-circ-zprop-5}, testing the validity of \eqref{eq_gen-fin-totW-scattering} for a generalised class of external plane-wave fields.

Let us first assume the initial electron at rest, i.e. $\pmb{q}_i = \pmb{0}$. We achieve this limit by setting $\lambda \rightarrow 1$ (cf. Sec.~\ref{sec_compton_nc}). In the considered POCS-$\text{CMS}_{n,cs}$ frame, this leads immediately to $n \omega \approx 0$ for arbitrary large harmonics which manifests clearly the soft nature of the corresponding field photons. Consequently, one can conclude for the corresponding upper limit in SPL
\eqn{
u_{n,\mathrm{coll}} \longrightarrow u_{n,\mathrm{soft}} \equiv \frac{2 n \omega}{m_{\star}} \ll 1,
}
where the term $n^2 [\omega/m]^2$ has been avoided due to its negligible smallness for all harmonic modes in the series. 
Hence, $u_{cs}$ will have small values during the remaining one-fold phase-space integral. 
A direct consequence of this is the following relation $k \cdot q_i \approx k \cdot q_f$.
Note that accordingly the trace from \eqref{eq_tr-result} will be less complicated, namely.:
\eqn{
\mathfrak{S}_{\mathrm{soft}} \approx 8 m^2 
- 2 e^2  \lL  \frac{ 1 + [1 + u_{cs}]^2 }{1 + u_{cs}}  \rL  \lL \gA_\varphi - \gA_\vartheta  \rL^2.
\label{eq_tr-spl-result}
}
Using \eqref{eq_wf-circ-zprop},
the final scattered photon with momentum $k_f$ can also be supposed as soft,
such that we conclude
\eqn{
\omega_f 
\approx  0.
\label{eq_spl-wf}
}
Since $u_{n,\mathrm{soft}} \ll 1$, we deduce in \eqref{eq_energy-ratio}
accordingly $q_{f,0} \gg \omega_f$ and hence $\omega_f \ll m_{\star}$. 
Applying \eqref{eq_spl-wf}, we may continue with another useful approximation:
\eqn{
q_{i,0} q_{f,0} 
\approx m_{\star}^2.
\label{eq_spl-qiqf}
}
Finally, after discussing briefly the soft nature of the emitted photon, whereas $u_{n,\mathrm{soft}}$ turned out to be negligible small
such that
\eqn{
[1 + u_{n,\mathrm{coll}}] \longrightarrow [1 + u_{n,\mathrm{soft}}] \approx 1
}
can be assumed, we proceed based on Eq.~\eqref{eq_Z-ncz-final} with the following argument 
\eqn{
Z_{n,\mathrm{coll}} \longrightarrow 
Z_{n,\mathrm{soft}}  \approx
2 n \frac{\xi}{\sqrt{1 + \xi^2}}
\sqrt{ \frac{u_{cs}}{ u_{n,\mathrm{soft}} } - \frac{u_{cs}^2}{u_{n,\mathrm{soft} }^2  } },
\label{eq_Z-spl-final-gen}
}
which is in full agreement with the literature, cf. e.g. Ref.~\cite{heinzl-10}\footnote{Note the sign change due to different convention for strength parameter, i.e. $\xi = \frac{- e^2 \mathfrak{a}^2}{m^2} \rightarrow \xi = \frac{e^2 \mathfrak{a}^2}{m^2}$.}.
Furthermore, following the SPL approximations for the functions defined in Sec.~\ref{sec_compton_nc} applies:
\begin{align}
\mathfrak{R}_{21,\mathrm{coll}} \longrightarrow \mathfrak{R}_{21,\mathrm{soft}} &\equiv 0, \label{eq_spl-beta21}\\
\mathfrak{T}_{31,\mathrm{coll}} \longrightarrow \mathfrak{T}_{31,\mathrm{soft}} &\equiv 8 m^2, \label{eq_spl-tau01}\\
\mathfrak{T}_{32,\mathrm{coll}} \longrightarrow \mathfrak{T}_{32,\mathrm{soft}} &\equiv 8 m^2 \xi^2   \lL \frac{1 + [1+u_{cs}]^2}{1+u_{cs}} \rL \label{eq_spl-tau32}.
\end{align}
Please note that each of the modes in the series correspond to a discretised momentum conservation law $n k^\mu + q_i^\mu = q_f^\mu + k_f^\mu$.
Multiplying with $q_i^\mu$ and using Eq.~\eqref{eq_spl-qiqf}  for the initial rest frame, one obtains $n \omega = \omega_f$.
The conservation is only satisfied if $n \geq 1$.  
Therefore, by applying all derivations \eqref{eq_Z-spl-final-gen} -  \eqref{eq_spl-tau32} to Eq.~\eqref{eq_totW-circ-zprop-5} we obtain the total Compton scattering probability:
\begin{framed}
\eqn{
W_{cs,\mathrm{soft}} = \frac{e^2    [2 \pi]   }{2^3  [2 q_{i,0}]  }
\sum_{n = 1}^{\infty} \int_{0}^{u_{n,\mathrm{soft}}} du_{cs}\ \hat{\mathfrak{J}}_{n,\mathrm{soft}}(Z_{n,\mathrm{soft}}(u_{cs}) ).
\label{eq_totW-spl-5}
}
\end{framed}
The non-trivial integrand in SPL reads as:
\eqnsplit{
\hat{\mathfrak{J}}_{n,\mathrm{soft}}(Z_{n,\mathrm{soft}}) &\equiv 
\frac{1}{[1+u_{cs}]^2} \bigg[
- \frac{  \mathfrak{T}_{31,\mathrm{soft}} }{2}  \mathcal{J}_n^2
+ \frac{  \mathfrak{T}_{32,\mathrm{soft}} }{8} \lL \mathcal{J}_{n-1}^2 + \mathcal{J}_{n+1}^2  - 2 \mathcal{J}_n^2  \rL
\bigg],
\label{eq_integrand-spl}
}
where $\mathcal{J}_j^2 \equiv \mathcal{J}_j^2(Z_{n,\mathrm{soft}})$. 
This result, often presented in the literature, is --- as we have recently shown and discussed in detail --- strongly restricted to the case of soft photons.

\end{appendix}

\newpage
\bibliographystyle{science}
\bibliography{akal-nonlinear2016-bibtex}

\end{document}